\documentclass[10pt,twocolumn,showpacs,longbibliography,amsmath,amssymb,osajnl,floatfix,superscriptaddress]{revtex4-1}
\usepackage[utf8]{inputenc}
\usepackage{graphicx} 
\usepackage{dcolumn} 
\usepackage{bm}
\usepackage{color}
\usepackage{txfonts}
\usepackage{microtype}
\usepackage{hyperref}
\usepackage[english]{babel}
\usepackage{siunitx}
\usepackage{url}
\usepackage{slashed}
\usepackage{gensymb}
\usepackage{epsfig}
\usepackage[normalem]{ulem}
\usepackage{amsmath} 
\usepackage{array}
\usepackage{booktabs}
\allowdisplaybreaks[4]

\UseRawInputEncoding

\begin{document}
\renewcommand{\figurename}{FIG}	
	
\title{Generation and Acceleration of Isolated-Attosecond Electron Bunch in a Hollow-Channel Plasma Wakefield}
	
\author{Liang-Qi Zhang}
\affiliation{School of Science, Shenzhen Campus of Sun Yat-sen University, Shenzhen 518107, People's Republic of China}
\author{Mei-Yu Si}
\email{simy@mail.sysu.edu.cn}
\affiliation{School of Science, Shenzhen Campus of Sun Yat-sen University, Shenzhen 518107, People's Republic of China}
\author{Tong-Pu Yu}
\email{tongpu@nudt.edu.cn}
\affiliation{Department of Physics, National University of Defense Technology, Changsha 410073, People's Republic of China}
\author{Yuan-Jie Bi}
\affiliation{School of Science, Shenzhen Campus of Sun Yat-sen University, Shenzhen 518107, People's Republic of China}
\author{Yong-Sheng Huang}
\email{huangysh59@mail.sysu.edu.cn}
\affiliation{School of Science, Shenzhen Campus of Sun Yat-sen University, Shenzhen 518107, People's Republic of China}	
	
\date{\today}
	
\begin{abstract}
 
We propose a novel scheme for generating and accelerating simultaneously a dozen-GeV isolated attosecond electron bunch from an electron beam-driven hollow-channel plasma target.
During the beam-target interaction, transverse oscillations of plasma electrons are induced, and subsequently, a radiative wakefield is generated. Meanwhile, a large number of plasma electrons of close to the speed of light are injected transversely from the position of the weaker radiative wakefield (e.g., the half-periodic node of the radiative wakefield) and converge towards the center of the hollow channel, forming an isolated attosecond electron bunch. Then, the attosecond electron bunch is significantly accelerated to high energies by the radiative wakefield.
It is demonstrated theoretically and numerically that this scheme can efficiently generate an isolated attosecond electron bunch with a charge of more than 2 nC, a peak energy up to 13 GeV of more than 2 times that of the driving electron beam, a peak divergence angle of less than 5 mmrad, a duration of 276 as, and an energy conversion efficiency of 36.7$\%$ as well as a high stability as compared with the laser-beam drive case.
Such an isolated attosecond electron bunch in the range of GeV would provide critical applications in ultrafast physics and high energy physics, etc.
	
\end{abstract}
	
\maketitle

Relativistic attosecond electron bunches have broad application prospects in ultrafast physics \cite{nobel2010Real,nobelPaul2001ObservationOA,nobelAnnePhysRevLett1996}, novel radiation sources \cite{RevModPhys.81.163,Sansone2011,Sansone2011,LiJX2015_PRL_attosecond_gamma} and free-electron laser injection \cite{undulatorGHAITH20211, undulatorDuris2019TunableIA, undulatorPhysRevLett.119.154801,xuUltracompactAttosecondXray2023}, etc.
For instances, attosecond electron bunches are used to detect ultrafast variation at the atomic scale in real time by ultrafast electron diffraction and microscopy in the field of ultrafast physics\cite{morimoto_diffraction_2018, nabben_attosecond_2023}; attosecond X/$\gamma$-rays can be generated by head-to-head collisions between an ultra-intense ultrashort laser and an attosecond electron bunch\cite{hu_attosecond_2021,zhao_all-optical_2022,li_ultra-bright_2017,zhuCollimatedGeVAttosecond2019,zhang_generation_2022,huRotatingAttosecondElectron2024}; and in the area of free-electron laser, attosecond electron bunches can be injected into undulators or wigglers to produce ultra-short brilliant soft/hard X-rays via self-amplified spontaneous emission or high-harmonic radiation\cite{huang_features_2021,petrillo_ultrahigh_2008,xu_generation_2022,fengBunchingEnhancementCoherent2024}.
Over the past few decades, a variety of ways of generating attosecond electron bunches have been attempted\cite{hu_attosecond_2021,zhao_all-optical_2022,li_ultra-bright_2017,zhuCollimatedGeVAttosecond2019,zhang_generation_2022,huRotatingAttosecondElectron2024,zhang_generation_2024,yu_bright_2024,Jiang2021DirectAO,liangHighrepetitionrateFewattosecondHighquality2020}.
Laser wakefield acceleration technique is one of the state-of-the-art candidates for generating high-quality attosecond electron bunches\cite{li_dense_2013,zhu_generation_2021,sun_generation_2024,deng_generation_2023,weikum_generation_2016,dodinStochasticExtractionPeriodic2007}.
Meanwhile, the interactions of lasers with solid-density plasma and near-critical-density plasma targets of different shapes and profiles have been explored as a method to generate attosecond electron bunches\cite{hu_dense_2018,hu_attosecond_2018,naumova_efficient_2006,naumovaAttosecondElectronBunches2004,kozakAllOpticalSchemeGeneration2019,liseykinaRelativisticAttosecondElectron2010,luttikhofGeneratingUltrarelativisticAttosecond2010}.
However, these attosecond electron bunches achieved are limited to either energy of MeV level\cite{hu_attosecond_2021,huRotatingAttosecondElectron2024,zhang_generation_2024,zhu_generation_2021}, high divergence\cite{zhang_generation_2022} or non-isolated bunches\cite{li_ultra-bright_2017}, so that they are not suitable for using in single-shot ultrafast diagnostics and imaging of materials.
Additionally, these methods of generating attosecond electrons are very sensitive to the spatiotemporal alignment accuracy between laser pulses and targets, and the energy conversion efficiency from laser to electron beams is only a few percent.

Currently, a promising way of producing and accelerating ultrashort and high-energy electron bunches is realized by means of an electron/proton beam-driven wakefield in under-dense plasma\cite{chen_acceleration_1985,katsouleas_physical_1986,lu_generating_2007,litos_high-efficiency_2014,li_ultrabright_2022,pompili_beam_2016,xu_high_2017,liangAccelerationElectronBunch2022}.
As an example, Xu $\textit{et al.}$ showed that under appropriate conditions, density downramp injection in a 3D blowout regime can generate brilliant electron beams with low energy spread\cite{xu_high_2017}.
However, the generated ultrashort electron bunches generally have a significant divergence angle and are absent of isolated attosecond duration. 
Especially, Si $\textit{et al.}$ numerically generated continuously high-intensity, long wavelength few-cycle infrared radiation through oscillations of a surface electron film driven by a relativistic electron beam in a dense plasma micro-tube\cite{si_relativistic-guided_2023}. The achieved infrared radiation field with uniform periodic distribution can be used to accelerate charged particles\cite{si_stable_2024}. 
However, in the scenario, the charged particles accelerated are externally injected and are not of an attosecond structure.
Most recently, Farmer $\textit{et al.}$ proposed a new scheme that plasma wakefield accelerators with higher energy can be periodically produced by a train of drive bunches driving an under-dense plasma \cite{farmerWakefieldRegenerationPlasma2024}.
However, only the acceleration of the injected electrons have been considered.
It still is a great challenge for the previously proposed methods to generate and accelerate simultaneously an isolated attosecond electron bunch with both the energy of dozen-GeV and the charge of a few nC.
Studying this key problem will help to gain insight into new mechanisms of beam-target interactions and provide an effective mechanism for realizing an isolated, high-energy and high-charge attosecond electron source.

\begin{figure}	[t]	 
	\setlength{\abovecaptionskip}{-0.2cm}
	\setlength{\belowcaptionskip}{-0.3cm}
	\centering
	\includegraphics[width=1.0\linewidth]{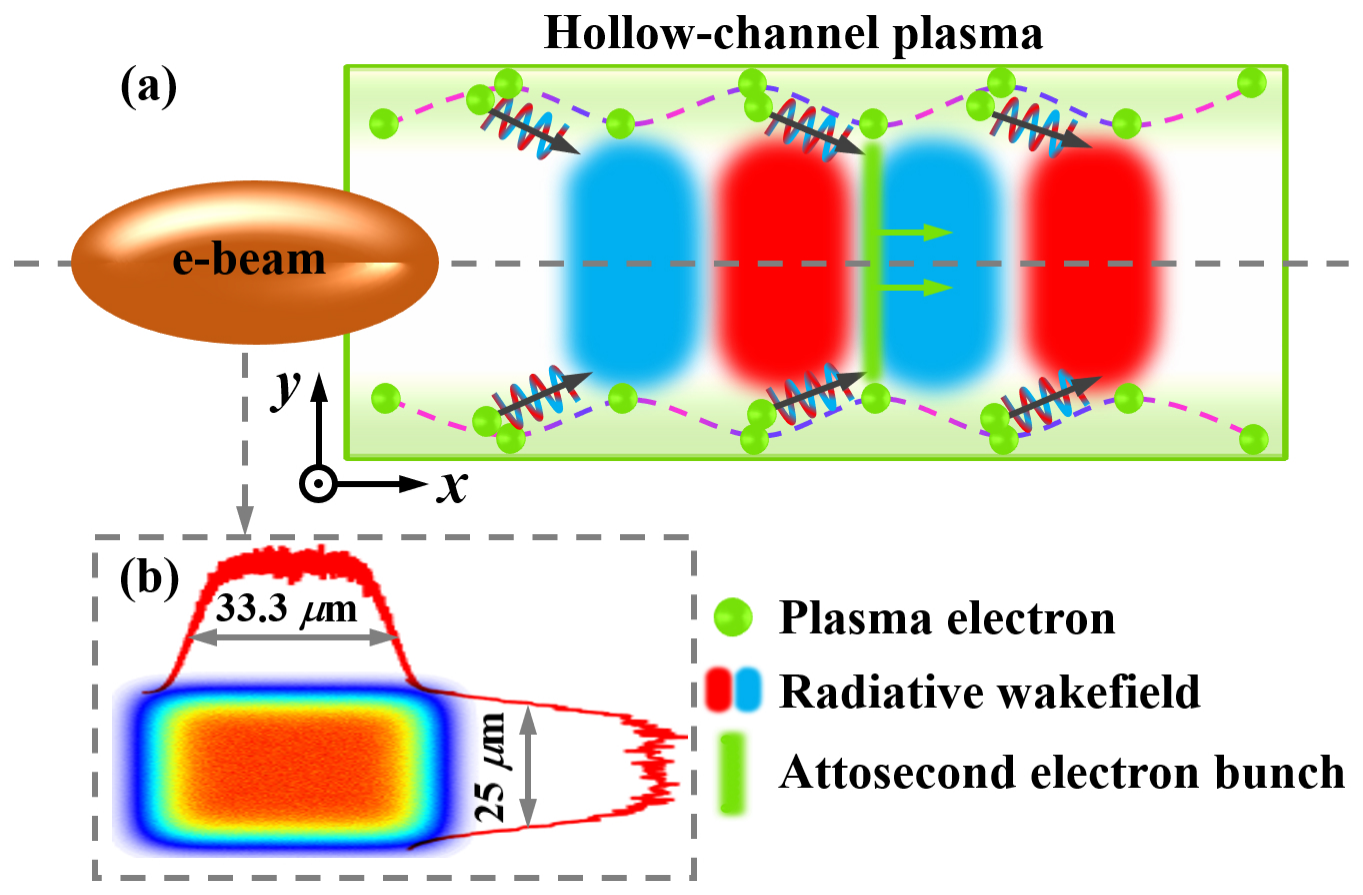}
	\vspace{10pt}
	\caption{(a) Schematic diagram for the generation and acceleration of isolated attosecond electron bunch. (b) Spatial profile of the driving electron beam in the $x-y$ plane.}
	\label{FIG.1}
\end{figure}

In this letter, we propose a novel scheme to efficiently generate and accelerate an isolated attosecond electron bunch with high-energy and high charge by using a relativistic electron beam propagating in a hollow-channel plasma target. 
As schematically shown in figure \ref{FIG.1}(a), when a driving electron beam with a reachable energy of 6 GeV and a charge of 5.78 nC in beam factories like the Facility for Advanced Accelerator Experimental Tests II (FACET-II) \cite{quSignatureCollectivePlasma2021,yakimenko_facet-ii_2019} passes through a hollow-channel plasma target, the electrostatic field of the driving electron beam perturbs the plasma electrons at the inner wall of the plasma channel, generating a transverse charge-separation field, and the plasma electrons at the inner wall oscillate collectively and transversally in the presence of the electrostatic field and the charge-separation field, giving rise to the generation of a radiative wakefield. 
Simultaneously, a large number of plasma electrons of close to the speed of light are transversely self-injected, at the locations where the radiative wakefield is weaker, into the hollow channel to form an isolated attosecond electron bunch and are constantly accelerated by the stable and high accelerator gradient radiative wakefield.
Ultimately, our scheme achieves an isolated attosecond electron bunch with a density up to 10 times the original plasma density, a charge of more than 2 nC, a peak energy of 13 GeV, a peak divergence angle of less than 5 mmrad, a duration of 276 as, and an energy conversion rate of 36.7$\%$. 
Furthermore, we found that the intensity of the radiative wakefield and the average energy of the isolated attosecond electron bunch can be well controlled by adjusting the spread and charge of the driving electron beam.

We perform two-dimensional particle-in-cell (2D-PIC) simulations with the open-source code EPOCH\cite{arber_contemporary_2015}.
The size of the simulation box is
$x\times y=100\,\mu{\mathrm{m}}\times60\,\mu{\mathrm{m}}$ 
with the spatial resolution 
$\Delta x\times\Delta y=0.0125\,\mu{\mathrm{m}}\times0.1\,\mu{\mathrm{m}}$, where the macro-particles per cell for the beam electrons, plasma electrons, and ions are 24, 12, and 8, respectively. 
A moving window is employed, which starts moving along the $x$ direction with the speed of light \textit{c} at 
$\textit{t}={\mathrm{100\,fs}}$. A relativistic electron beam with an energy of 6 GeV and a charge of 5.78 nC enters the simulation box from the left boundary and propagates along the +$x$ direction.
As shown in figure \ref{FIG.1}(b), its spatial profile is characterized by a super-Gaussian density profile of
$n_{b}=n_{b0}\mathrm{exp}\left(-\frac{(x-x_{0})^{6}}{\sigma_{x}^{6}}-\frac{y^{6}}{\sigma_{y}^{6}}\right)$ in both longitudinal and transverse direction, where $n_{b0}=1.4\times10^{19}\,\mathrm{cm}^{-3}$, $x_0=16.65\,\mu{\mathrm{m}}$, $\sigma_{x}=10\,\mu{\mathrm{m}}$ (corresponding to full width at half
maximum (FWHM) of $33.3\,\mu{\mathrm{m}}$), $\sigma_{y}=7.5\,\mu{\mathrm{m}}$ (corresponding to FWHM of $25.0\,\mu{\mathrm{m}}$), respectively.
It is worth noting that an electron beam with 10 GeV energy and $3\times10^{19}\,\mathrm{cm}^{-3}$ peak density represents the state of the art at the FACET-II \cite{quSignatureCollectivePlasma2021,yakimenko_facet-ii_2019}, which is very close to the driving electron beam parameters (i.e., energy of 6 GeV and peak density of $1.4\times10^{19}\,\mathrm{cm}^{-3}$) of our scheme. 
Apart from that, higher-charge ($\sim$100 nC) electron beams will be achieved in the Argonne Wakefield Accelerator (AWA) in the near future\cite{clarkeUSAdvancedNovel2022}.
The hollow-channel plasma target can be viewed as a hollow, long, straight carbon tube with the wall made of fully ionized plasma, while the interior is vacuum. In our scheme, the carbon tube is located between $x=0\,\mu{\mathrm{m}}$ and $15000\,\mu{\mathrm{m}}$, whose inner and outer radius are $20\,\mu{\mathrm{m}}$ and $25\,\mu{\mathrm{m}}$, respectively. The plasma electron density is
$n_{p}=1.1\times10^{20}\,\mathrm{cm}^{-3}$.
Such a plasma channel can be created by laser-irradiated carbon nanotubes\cite{ostling_electronic_1997,wang_interactions_2004,martin-luna_excitation_2023} or laser near-critical density gas interaction\cite{ospina-bohorquez_laser-driven_2023}.
In our simulations, the absorbing boundary condition is utilized for both fields and particles in the transverse directions.

\begin{figure}[h]      
	\centering
	\includegraphics[width=1.0\linewidth]{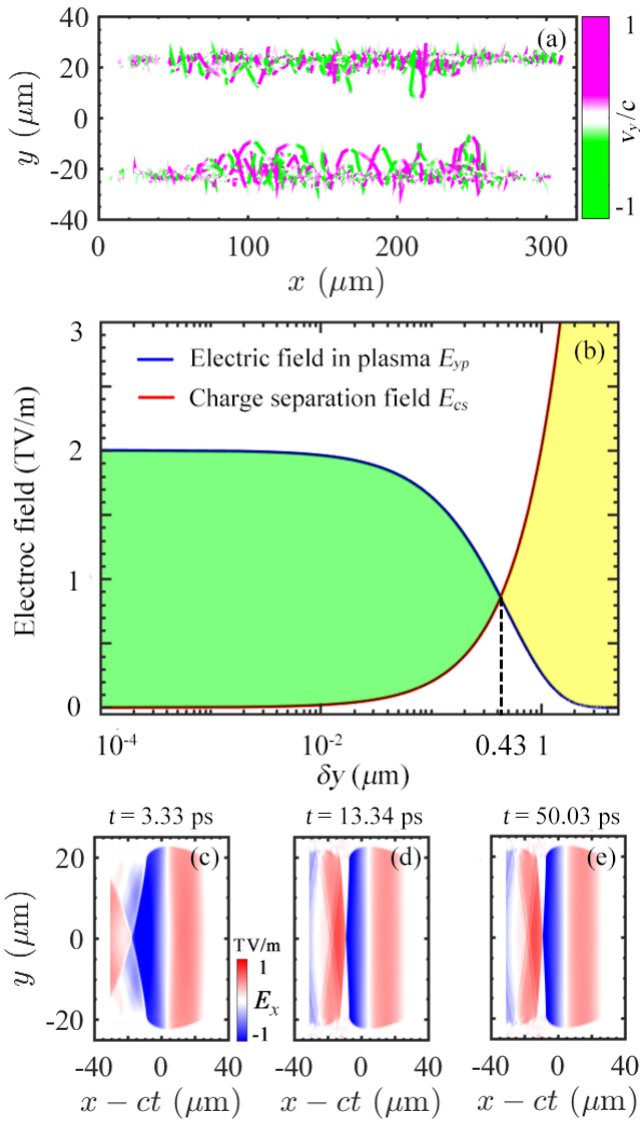}
	\caption{(a) Trajectories of some typical plasma electrons with transverse different velocity $v_y$. (b) The transverse electric field $E_{y p}\,=\,E_{0}e^{-\sqrt{k_{x}^{2}-1+\omega_{p e}^{2}}\delta y}$ and the charge-separation field $E_{cs}=en_p{\delta y}/\epsilon_o$ as a function of the transverse oscillation distance of plasma electrons $\delta y$. (c-e) Longitudinal mid-infrared radiation fields $E_x$ at $t = {\mathrm{3.33\,ps}}$, ${\mathrm{13.34\,ps}}$, and ${\mathrm{50.03\,ps}}$, respectively.}
	\label{FIG.2}
\end{figure}

Figure \ref{FIG.2} illustrates the generation and propagation process of the radiative wakefield. Figure\ref{FIG.2}(a) shows the typical trajectories of plasma electrons from the inner walls. It can be seen that the plasma electrons on the inner wall of the hollow channel are collectively oscillating transversely at about 0.83$c$. This is because when the driving electron beam passes through the hollow channel, the Coulomb repulsive force forms in the transverse direction, which excites the plasma electrons in the inner wall, kicking most of the plasma electrons in the skinning depth towards the outer wall by the transverse electric field force $E_{y p}\,=\,E_{0}e^{-\sqrt{k_{x}^{2}-1+\omega_{p e}^{2}}\delta y}$, where $E_0$ is the initial transverse electric field of the driving electron, $k_x$ is the wavenumber in the $x$-direction, $\omega_{p e}=\sqrt{e^2n_p/(m_e\varepsilon_0)}$ is the plasma frequency, $e$ is the charge of the electron, $m_e$ is the mass of the electron, $\varepsilon_0$ is the vacuum permittivity, and $\delta y$ is the transverse oscillation distance of the plasma electrons. 
Meanwhile, a charge-separation field $E_{cs}=en_p{\delta y}/\epsilon_o$
is formed between the kicked plasma electrons and the near carbon ions, so that the plasma electrons kicked towards the outer wall return back to the skin layer. 
Figure\ref{FIG.2}(b) shows the evolution of the transverse electric field and the charge-separation field as a function of $\delta y$. 
One can see that the charge-separation field keeps a rising trend while the transverse electric field slows a falling trend with increasing $\delta y$. When $\delta y$ is 0.43${\ }\mu$m, the charge-separation field and the transverse electric field are balanced, so the plasma electrons attain the maximum momentum here. Thereafter, the charge-separation field becomes larger than the transverse electric field, and the plasma electron is pulled back.

As the plasma electrons oscillate transversely under the combined force of transverse electric field force and charge-separation force, the net energy of the plasma electrons can be obtained by
$$e\int_{0}^{\delta_{y}}(E_{yp}-E_{cs})d y=(1/\sqrt{1-\beta_{p}}-1)m_{e}c^{2},\eqno{(1)}$$
where $\beta_{p}=v_y/c$, and $v_y$ is the transverse speed of plasma electrons.
When the energetic plasma electrons undergo collective oscillations, giving rise to the generation of high-gradient and stable radiative wakefields, the generated radiative wakefields can be estimated by\cite{Jackson1998}
$$E_{rad}=\frac{e}{4\pi\varepsilon_{0}}[\frac{\vec{n}\times(\vec{n}\times\dot{\vec{\beta_{p}}})}{c(1-\beta_{p}cos\theta)^{3}R}], \eqno{(2)}$$
where $\dot{\vec{\beta_{p}}}=\frac{e\vec{E}}{m_ec\gamma_p^3}$, $\vec{E}={\vec{E}}_{yp}+{\vec{E}}_{cs}$, $\theta = \langle \vec{n}, \vec{\beta_{p}} \rangle$, $R$ is the distance from the reference point to the inner wall of the hollow plasma channel,  $\gamma_p$ is the Lorentz factor of the plasma electron.
Integrating the above equation, we finally obtain total longitudinal radiative wakefields:
$$E_{rad,x}^{total}=\int_{0}^{\theta}E_{rad,x} \mathrm{d}\theta=\frac{e\dot{\vec{\beta_{p}}}n_p\delta_{y}y}{2\varepsilon_{0}c}
\int_{0}^{\theta}\frac{\sin\theta}{\left(1-\beta_{p}\mathrm{cos}\theta\right)^{3}}\mathrm{d}\theta, \eqno{(3)}$$
where $E_{rad,x}=E_{rad}\sin\theta$ is the longitudinal radiative wakefield, $y=R\cos\theta$ is the projection of $R$ in the $y$-direction.
For example, taking the transverse electric field $E=1.35$ TV/m, the transverse oscillation distance of plasma electrons $\delta y=0.43{\ }\mu$m, the distance from the reference point to the inner wall of the hollow plasma channel $y=20{\ }\mu$m, normalized velocity $\beta_{p}=0.83$, we can estimate the total radiative wakefields $E_{rad,x}^{total}=1.02$ TV/m, which is in consistent with the numerical result of $E_{x,max}=1$ TV/m, as shown in Fig. \ref{FIG.2}[(c)-(e)]. 
Such a stable and ultra-high gradient radiative wakefield can be used to accelerate the electrons, positrons, protons and even muons, etc, to extremely high energies\cite{yu_bright_2024}.

\begin{figure}[h]
	\centering
	\includegraphics[width=1.0\linewidth]{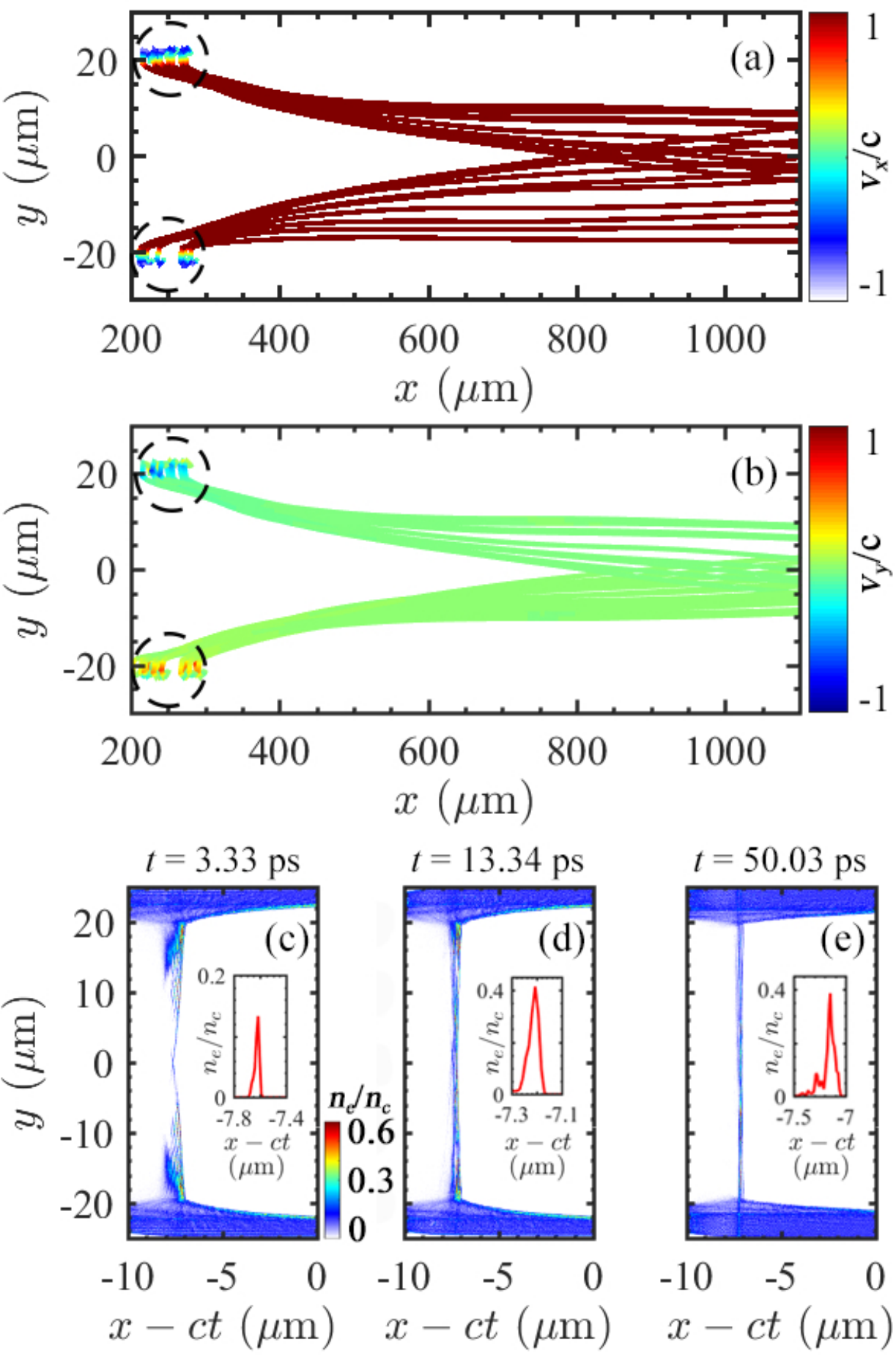}
	\caption{(a) Trajectories of attosecond electron bunch with $v_y$. 
		(b) Trajectories of attosecond electron bunch with $v_x$. (c-e) The corresponding plasma electron number densities $n_e / n_c$. Insets are the generated attosecond pulse with ${\mathrm{114\,as}}$, ${\mathrm{150\,as}}$ and ${\mathrm{276\,as}}$, respectively. Here, the plasma electron number density are normalized by $n_c=1.1\times10^{27}\,\mathrm{m}^{-3}$.}
	\label{FIG.3}
\end{figure}

When the plasma electrons on the inner wall of the hollow channel oscillate back and forth transversely, some of them close to the speed of light eject transversely from the position of weaker radiative wakefield, (e.g., the half-periodic node of the radiative wakefield) and converge towards the center of the hollow channel to form an isolated attosecond electron bunch.
Figure \ref{FIG.3}(a) shows the trajectories and longitudinal velocity distributions of the attosecond electrons in the ($x, y$) phase space. It can be seen that the longitudinal velocity of the attosecond electrons is close to the speed of light and maintains good collimation in the longitudinal direction. 
Figure \ref{FIG.3}(b) shows the trajectory and transversal velocity distribution of the attosecond electrons in the same phase space. The transversal velocity of the attosecond electrons is much less than the speed of light, resulting in a low divergence angle of the attosecond electron bunch.
Figures \ref{FIG.3}(c)-(e) present the evolution of electron number density distributions of the attosecond electrons bunch at $t = {\mathrm{3.33\,ps}}$, ${\mathrm{13.34\,ps}}$ and ${\mathrm{50.03\,ps}}$, respectively.
The electrons escaping from the inner wall of the hollow channel are firstly accelerated and then decelerated (as indicated by the dashed circle in Figure \ref{FIG.3}(a)) in the $+y$ direction at $t = {\mathrm{3.33\,ps}}$, and most of these electrons with densities of $0.3\sim0.6\,n_c$ are mainly concentrated in the region of $-20\,\mu{\mathrm{m}}$\textless$y$\textless$-10\,\mu{\mathrm{m}}$ and $10\,\mu{\mathrm{m}}$\textless$y$\textless$20\,\mu{\mathrm{m}}$, and only a few electrons with densities of $\leq{0.2n_c}$ are distributed in the $-10\,\mu{\mathrm{m}}$\textless$y$\textless$10\,\mu{\mathrm{m}}$ region, forming an isolated attosecond electron bunch, as shown in the Figures \ref{FIG.3}(c)-(e).
Since the transverse velocity of the electrons of the attosecond electron bunch is much smaller than their longitudinal velocity, the electrons are predominately accelerated longitudinally by the radiative wakefield along the $x$-direction and simultaneously focused in the transverse direction. 
Most of the electrons in the $-20\,\mu{\mathrm{m}}$\textless$y$\textless$-10\,\mu{\mathrm{m}}$ and $10\,\mu{\mathrm{m}}$\textless$y$\textless$20\,\mu{\mathrm{m}}$ regions move transversely to the $-10\,\mu{\mathrm{m}}$\textless$y$\textless$10\,\mu{\mathrm{m}}$ region at $t = {\mathrm{13.34\,ps}}$ and ${\mathrm{13.34\,ps}}$, causing the peak number density of $0.4\,n_c$, as shown in the figures \ref{FIG.3}(d) and \ref{FIG.3}(e). 
However, this enlarges the duration of the attosecond electron bunch from 114 as to 276 as at $t = {\mathrm{3.33\,ps}}$ and $t = {\mathrm{50.03\,ps}}$, respectively, as shown in the inset of figures \ref{FIG.3}(c)-(e).
In addition, the charge of the attosecond electron bunch decreases from 5.3 nC at $t = {\mathrm{3.33\,ps}}$ to 2.4 nC at $t = {\mathrm{50.03\,ps}}$, which indicates that a large number of the electrons are dispersed during acceleration from the radiative wakefield axis.

\begin{figure}[h]
	\centering
	\includegraphics[width=1.0\linewidth]{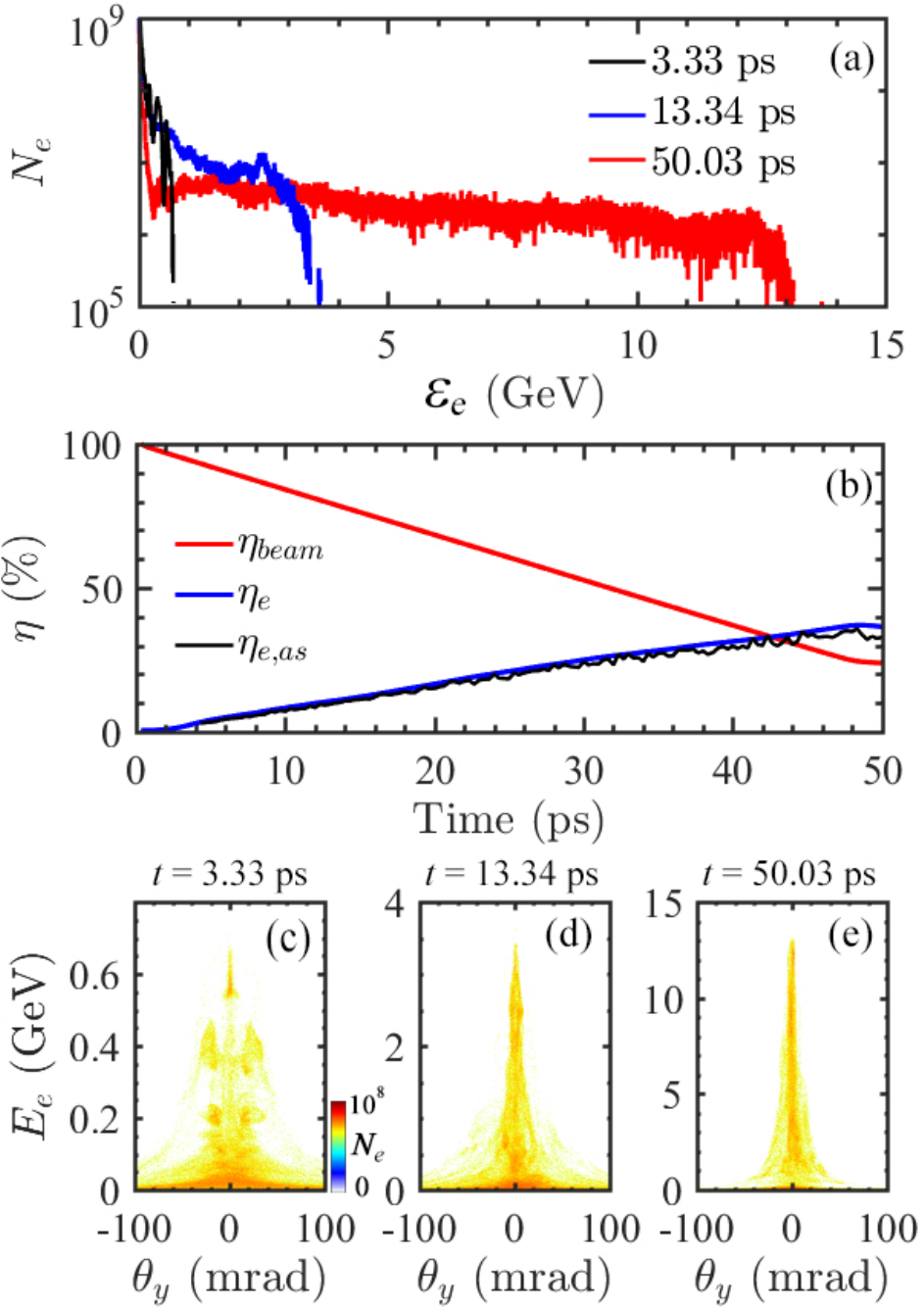}
	\caption{(a) Energy spectra of electrons at $t = {\mathrm{3.33\,ps}}$, $t = {\mathrm{13.34\,ps}}$ and $t = {\mathrm{50.03\,ps}}$. (b) Red, blue and black curves represent energy conversion efficiency of driving electron beam $\eta_{beam}$, total plasma electron $\eta_{e}$ and attosecond electron bunch $\eta_{e,as}$ as a function of time, respectively. (c-e) Angular distribution of attosecond electron energy at $t = {\mathrm{3.33\,ps}}$, ${\mathrm{13.34\,ps}}$, and ${\mathrm{50.03\,ps}}$, respectively.}
	\label{FIG.4}
\end{figure}

The attosecond electrons ejected into the hollow channel are continuously accelerated by the stable radiative wakefield. 
Figure \ref{FIG.4}(a) shows the energy spectra of electrons at $t = {\mathrm{3.33\,ps}}$, ${\mathrm{13.34\,ps}}$, and ${\mathrm{50.03\,ps}}$, respectively. The cut-off energy of the attosecond electron bunch is up to 600 MeV at $t = {\mathrm{3.33\,ps}}$.
They are accelerated by the radiative wakefield of $E_{x,max}=1$ TV/m for a time duration of ${\Delta}t=$10.01 ps. According to
$\varepsilon_{e,max} \sim eE_{x,max}l_0$,
where $l_0=c{\Delta}t$ is the accelerating distance, we may estimate simply the maximum energy of attosecond electrons $\varepsilon_{e,max}$ is close to 3.33 GeV at $t = {\mathrm{13.34\,ps}}$, which is excellent agreement with the simulation results, as shown in figure \ref{FIG.4}(a).
The attosecond electrons are further accelerated to the maximum energy of 13 GeV at $t = {\mathrm{50.03\,ps}}$.
It is noted that the energy distribution of the attosecond electron bunch is broad. This is due to the fact that the injected electrons are mainly distributed at the nodes of the longitudinal radiative wakefield half-cycle, where the intensity of the negative electric field is quasi-linearly distributed in the $x$ direction. Electrons at the head of the attosecond bunch are accelerated at a higher radiative wakefield, while the electrons at the tail of that are accelerated at a lower radiative wakefield. This also makes the attosecond electron energy conversion efficiency decrease linearly with respect to accelerating time, as shown in figure \ref{FIG.4}(b). In addition, the energy conversion efficiency from the driving electron beam to the attosecond electron bunch is as high as 36.7$\%$, which is ten times higher than achieved by other plasma-based mechanisms \cite{huRotatingAttosecondElectron2024,zhu_generation_2021}.

Figures \ref{FIG.4}(c)-(e) present the energy angle distribution of the attosecond electron bunch.
At $t = {\mathrm{3.33\,ps}}$, the angular distribution of the electrons exhibits a `trident' shape due to the fact that most of the electrons with a large beam-divergence are distributed near the inner wall of the hollow channel, which is in agreement with the distribution of the number density of electrons as shown in Figure \ref{FIG.3}(a). While the electrons are accelerated by the radiative wakefield, they continuously move from both sides of the inner wall of the hollow channel to the center of that, exhibiting a narrower divergence angle in the $y$ direction.
Finally, the minimum divergence angle of the attosecond electron bunch is nearly $5\,\text{mrad}$ for the electrons of cut-off energy of $13\,\text{GeV}$ at $t = {\mathrm{50.03\,ps}}$.

\begin{figure}[h]
	\centering
	\includegraphics[width=1.0\linewidth]{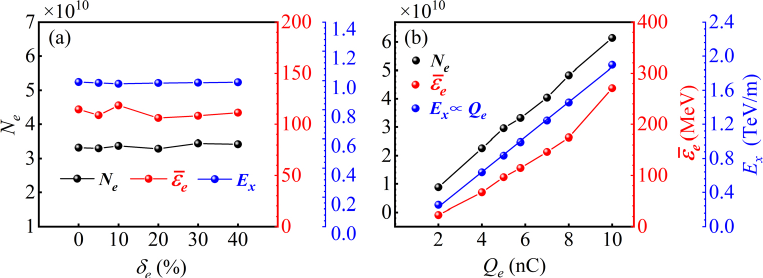}
	\caption{Yield $N_e$ and average energy $\overline{\varepsilon}_e$ of attosecond electrons and maximum acceleration intensity $E_{x}$ of radiative wakefield as a function of the driving electron beam energy spread $\delta_e$  (a) and charge $Q_e$ (b),  respectively.}
	\label{FIG.5}
\end{figure}

In order to demonstrate the robustness of the scheme, we perform additional simulations to investigate the effect of the driving electron beam energy spread $\delta_e$ and charge $Q_e$ on the average energy $\overline{\varepsilon}_e$ of attosecond electrons and maximum acceleration intensity $E_{x}$ of the radiative wakefield, as shown in Figure \ref{FIG.5}. Here, the maximum acceleration intensity $E_{x}$ of the radiative wakefield is measured at $t = {\mathrm{3.33\,ps}}$. First, we keep all other parameters unchanged but vary $\delta_e$ from 0 to 40$\%$.
Figure \ref{FIG.5}(a) presents that the maximum acceleration intensity $E_{x}$ of the radiative wakefield maintains $\sim\text{1\,TV/m}$ with respect to the increase of $\delta_e$, which indicates that the $E_{x}$ is insensitive to the $\delta_e$. 
One can see that yield $N_e$ of attosecond electrons remains near $3.36\times10^{10}$ (i.e., 5.38 nC) with regard to $\delta_e$, which indicates $N_e$ maintains an almost constant trend.
The higher electron average energy $\overline{\varepsilon}_e$ can be achieved by increasing $\delta_e$.
One can see that as $\delta_e$ changes from 0 to $10\%$, the electron average energy $\overline{\varepsilon}_e$ first decreases by 6 MeV and then increases by 9 MeV.
A 10$\%$ change of $\delta_e$ only gives rise to $\text{3\,MeV}$ increase of the optimal electron average energy $\overline{\varepsilon}_e$.
When $\delta_e$ changes from $20\%$ to $40\%$, the electron average energy $\overline{\varepsilon}_e$ increases from 106 MeV to 111 MeV. 
In our scheme, hence, the effect of $\delta_e$ of the driving electron beam on the yield $N_e$ and average energy $\overline{\varepsilon}_e$ of attosecond electrons can be ignored, facilitating future experimental demonstration.

Secondly, we keep $\delta_e$ = 0 but vary the driving electron beam charge $Q_e$ from $\text{2\,nC}$ to $\text{10\,nC}$.
It is shown in Figure \ref{FIG.5}(b) that the attosecond electron yield $N_e$, the attosecond electron average energy $\overline{\varepsilon}_e$ and the maximum acceleration intensity $E_{x}$ of the radiative wakefield increase with the driving electron beam charge $Q_e$. 
The increase of the driving electron beam charge $Q_e$ leads to an increase of the transverse electric field ${E}$, giving rise to stronger oscillation of plasma electrons in the inner wall of the hollow palsma channel. 
Thus, the maximum acceleration intensity $E_{x}$ of the radiative wakefield has a linear scale law ($E_x\propto{\dot{\vec{\beta_{p}}}}\propto{E}\propto{Q_e}$) with respect to the driving electron beam charge $Q_e$. This indicates that the radiative wakefield is non-coherent radiation field.
Meanwhile, when $Q_e$ = 2 nC, $\overline{\varepsilon}_e$ is $\text{23\,MeV}$ and $E_x$ is 0.25 TV/m. While the driving electron beam charge increases from $Q_e$ = 2 nC to $Q_e$ = 10 nC, $\overline{\varepsilon}_e$ and $E_x$ increase significantly by 12 times and seven times, respectively.

In summary, we here propose a novel and efficient scenario for the generation of an isolated attosecond (276 as) electron bunch. In this scenario, a single attosecond electron bunch with high-energy (13 GeV) and high-charge (more than 2 nC) is efficiently generated. The energy conversion rate reaches 36.7$\%$ from the driving beam to the isolated attosecond electron bunch. Furthermore, we discuss the effect of the energy spread and charge of the driving electron beam with respect to the energy of the attosecond electron bunch and the intensity of the radiated wakefield.
With such a dozen-GeV electron bunch, it is able to image matters through much thicker materials. 
With resolution of attosecond scale, it is well suited to study the evolution of fast material processes under dynamic conditions.
Such an isolated, high-energy, high-charge, and attosecond electron bunch in the range of GeV driven by an electron beam can be especially used for higher resolution of high surface density matter in imaging experiments of materials\cite{merrill_demonstration_2018,Walstrom:2015bbt} and thus to obtain a comprehensive and integrated performance of materials under extreme conditions\cite{bell_electron_2014_10.1117/12.2061952}.

\hspace*{\fill}

\textit{\textbf{Acknowledgement}}
This work was supported by Project of Gamma-Gamma Collider and Integrated Beam Flow Facility (Phase I) Validation Device at SYSU (2403-000000-05-03-714165), Research Project of SYSU (74140-71020003, 74140-71020006) and the National Natural Science Foundation of China (12375244, 12135009). 
The authors acknowledge Guangdong Provincial Key Laboratory of Gamma-Gamma Collider and Its Comprehensive
Applications, Guangdong Provincial Key Laboratory of Advanced Particle Detection Technology.
The authors also acknowledge the access to the EPOCH code developed by University of Warwick, and the High-performance Computing Public Platform (Shenzhen Campus) of SYSU for providing computing resources.

\bibliographystyle{apsrev4-1}
\bibliography{11}

\begin{thebibliography}{62}%
\makeatletter
\providecommand \@ifxundefined [1]{%
 \@ifx{#1\undefined}
}%
\providecommand \@ifnum [1]{%
 \ifnum #1\expandafter \@firstoftwo
 \else \expandafter \@secondoftwo
 \fi
}%
\providecommand \@ifx [1]{%
 \ifx #1\expandafter \@firstoftwo
 \else \expandafter \@secondoftwo
 \fi
}%
\providecommand \natexlab [1]{#1}%
\providecommand \enquote  [1]{``#1''}%
\providecommand \bibnamefont  [1]{#1}%
\providecommand \bibfnamefont [1]{#1}%
\providecommand \citenamefont [1]{#1}%
\providecommand \href@noop [0]{\@secondoftwo}%
\providecommand \href [0]{\begingroup \@sanitize@url \@href}%
\providecommand \@href[1]{\@@startlink{#1}\@@href}%
\providecommand \@@href[1]{\endgroup#1\@@endlink}%
\providecommand \@sanitize@url [0]{\catcode `\\12\catcode `\$12\catcode
  `\&12\catcode `\#12\catcode `\^12\catcode `\_12\catcode `\%12\relax}%
\providecommand \@@startlink[1]{}%
\providecommand \@@endlink[0]{}%
\providecommand \url  [0]{\begingroup\@sanitize@url \@url }%
\providecommand \@url [1]{\endgroup\@href {#1}{\urlprefix }}%
\providecommand \urlprefix  [0]{URL }%
\providecommand \Eprint [0]{\href }%
\providecommand \doibase [0]{http://dx.doi.org/}%
\providecommand \selectlanguage [0]{\@gobble}%
\providecommand \bibinfo  [0]{\@secondoftwo}%
\providecommand \bibfield  [0]{\@secondoftwo}%
\providecommand \translation [1]{[#1]}%
\providecommand \BibitemOpen [0]{}%
\providecommand \bibitemStop [0]{}%
\providecommand \bibitemNoStop [0]{.\EOS\space}%
\providecommand \EOS [0]{\spacefactor3000\relax}%
\providecommand \BibitemShut  [1]{\csname bibitem#1\endcsname}%
\let\auto@bib@innerbib\@empty
\bibitem [{\citenamefont {Goulielmakis}\ \emph {et~al.}(2010)\citenamefont
  {Goulielmakis}, \citenamefont {Loh}, \citenamefont {Wirth}, \citenamefont
  {Santra}, \citenamefont {Rohringer}, \citenamefont {Yakovlev}, \citenamefont
  {Zherebtsov}, \citenamefont {Pfeifer}, \citenamefont {Azzeer}, \citenamefont
  {Kling}, \citenamefont {Leone},\ and\ \citenamefont
  {Krausz}}]{nobel2010Real}%
  \BibitemOpen
  \bibfield  {author} {\bibinfo {author} {\bibfnamefont {E.}~\bibnamefont
  {Goulielmakis}}, \bibinfo {author} {\bibfnamefont {Z.-H.}\ \bibnamefont
  {Loh}}, \bibinfo {author} {\bibfnamefont {A.}~\bibnamefont {Wirth}}, \bibinfo
  {author} {\bibfnamefont {R.}~\bibnamefont {Santra}}, \bibinfo {author}
  {\bibfnamefont {N.}~\bibnamefont {Rohringer}}, \bibinfo {author}
  {\bibfnamefont {V.~S.}\ \bibnamefont {Yakovlev}}, \bibinfo {author}
  {\bibfnamefont {S.}~\bibnamefont {Zherebtsov}}, \bibinfo {author}
  {\bibfnamefont {T.}~\bibnamefont {Pfeifer}}, \bibinfo {author} {\bibfnamefont
  {A.~M.}\ \bibnamefont {Azzeer}}, \bibinfo {author} {\bibfnamefont {M.~F.}\
  \bibnamefont {Kling}}, \bibinfo {author} {\bibfnamefont {S.~R.}\ \bibnamefont
  {Leone}}, \ and\ \bibinfo {author} {\bibfnamefont {F.}~\bibnamefont
  {Krausz}},\ }\href {\doibase 10.1038/nature09212} {\bibfield  {journal}
  {\bibinfo  {journal} {Nature}\ }\textbf {\bibinfo {volume} {466}},\ \bibinfo
  {pages} {739} (\bibinfo {year} {2010})}\BibitemShut {NoStop}%
\bibitem [{\citenamefont {Paul}\ \emph {et~al.}(2001)\citenamefont {Paul},
  \citenamefont {Toma}, \citenamefont {Breger},\ and\ \citenamefont
  {Mullot}}]{nobelPaul2001ObservationOA}%
  \BibitemOpen
  \bibfield  {author} {\bibinfo {author} {\bibfnamefont {P.~M.}\ \bibnamefont
  {Paul}}, \bibinfo {author} {\bibfnamefont {E.~S.}\ \bibnamefont {Toma}},
  \bibinfo {author} {\bibfnamefont {P.}~\bibnamefont {Breger}}, \ and\ \bibinfo
  {author} {\bibfnamefont {G.~a.}\ \bibnamefont {Mullot}},\ }\href {\doibase
  10.1126/SCIENCE.1059413} {\bibfield  {journal} {\bibinfo  {journal}
  {Science}\ }\textbf {\bibinfo {volume} {292}},\ \bibinfo {pages} {1689 }
  (\bibinfo {year} {2001})}\BibitemShut {NoStop}%
\bibitem [{\citenamefont {Antoine}\ \emph {et~al.}(1996)\citenamefont
  {Antoine}, \citenamefont {L'Huillier},\ and\ \citenamefont
  {Lewenstein}}]{nobelAnnePhysRevLett1996}%
  \BibitemOpen
  \bibfield  {author} {\bibinfo {author} {\bibfnamefont {P.}~\bibnamefont
  {Antoine}}, \bibinfo {author} {\bibfnamefont {A.}~\bibnamefont {L'Huillier}},
  \ and\ \bibinfo {author} {\bibfnamefont {M.}~\bibnamefont {Lewenstein}},\
  }\href {\doibase 10.1103/PhysRevLett.77.1234} {\bibfield  {journal} {\bibinfo
   {journal} {Phys. Rev. Lett.}\ }\textbf {\bibinfo {volume} {77}},\ \bibinfo
  {pages} {1234} (\bibinfo {year} {1996})}\BibitemShut {NoStop}%
\bibitem [{\citenamefont {Krausz}\ and\ \citenamefont
  {Ivanov}(2009)}]{RevModPhys.81.163}%
  \BibitemOpen
  \bibfield  {author} {\bibinfo {author} {\bibfnamefont {F.}~\bibnamefont
  {Krausz}}\ and\ \bibinfo {author} {\bibfnamefont {M.}~\bibnamefont
  {Ivanov}},\ }\href {\doibase 10.1103/RevModPhys.81.163} {\bibfield  {journal}
  {\bibinfo  {journal} {Rev. Mod. Phys.}\ }\textbf {\bibinfo {volume} {81}},\
  \bibinfo {pages} {163} (\bibinfo {year} {2009})}\BibitemShut {NoStop}%
\bibitem [{\citenamefont {Sansone}\ \emph {et~al.}(2011)\citenamefont
  {Sansone}, \citenamefont {Poletto},\ and\ \citenamefont
  {Nisoli}}]{Sansone2011}%
  \BibitemOpen
  \bibfield  {author} {\bibinfo {author} {\bibfnamefont {G.}~\bibnamefont
  {Sansone}}, \bibinfo {author} {\bibfnamefont {L.}~\bibnamefont {Poletto}}, \
  and\ \bibinfo {author} {\bibfnamefont {M.}~\bibnamefont {Nisoli}},\ }\href
  {\doibase 10.1038/nphoton.2011.167} {\bibfield  {journal} {\bibinfo
  {journal} {Nature Photonics}\ }\textbf {\bibinfo {volume} {5}},\ \bibinfo
  {pages} {655} (\bibinfo {year} {2011})}\BibitemShut {NoStop}%
\bibitem [{\citenamefont {Li}\ \emph {et~al.}(2015)\citenamefont {Li},
  \citenamefont {Hatsagortsyan}, \citenamefont {Galow},\ and\ \citenamefont
  {Keitel}}]{LiJX2015_PRL_attosecond_gamma}%
  \BibitemOpen
  \bibfield  {author} {\bibinfo {author} {\bibfnamefont {J.-X.}\ \bibnamefont
  {Li}}, \bibinfo {author} {\bibfnamefont {K.~Z.}\ \bibnamefont
  {Hatsagortsyan}}, \bibinfo {author} {\bibfnamefont {B.~J.}\ \bibnamefont
  {Galow}}, \ and\ \bibinfo {author} {\bibfnamefont {C.~H.}\ \bibnamefont
  {Keitel}},\ }\href {\doibase 10.1103/PhysRevLett.115.204801} {\bibfield
  {journal} {\bibinfo  {journal} {Phys. Rev. Lett.}\ }\textbf {\bibinfo
  {volume} {115}},\ \bibinfo {pages} {204801} (\bibinfo {year}
  {2015})}\BibitemShut {NoStop}%
\bibitem [{\citenamefont {Ghaith}\ \emph {et~al.}(2021)\citenamefont {Ghaith},
  \citenamefont {Couprie}, \citenamefont {Oumbarek-Espinos}, \citenamefont
  {Andriyash}, \citenamefont {Massimo}, \citenamefont {Clarke}, \citenamefont
  {Courthold}, \citenamefont {Bayliss}, \citenamefont {Bernhard}, \citenamefont
  {Trunk}, \citenamefont {Valléau}, \citenamefont {Marcouillé}, \citenamefont
  {Chancé}, \citenamefont {Licciardi}, \citenamefont {Malka}, \citenamefont
  {Nguyen},\ and\ \citenamefont {Dattoli}}]{undulatorGHAITH20211}%
  \BibitemOpen
  \bibfield  {author} {\bibinfo {author} {\bibfnamefont {A.}~\bibnamefont
  {Ghaith}}, \bibinfo {author} {\bibfnamefont {M.-E.}\ \bibnamefont {Couprie}},
  \bibinfo {author} {\bibfnamefont {D.}~\bibnamefont {Oumbarek-Espinos}},
  \bibinfo {author} {\bibfnamefont {I.}~\bibnamefont {Andriyash}}, \bibinfo
  {author} {\bibfnamefont {F.}~\bibnamefont {Massimo}}, \bibinfo {author}
  {\bibfnamefont {J.}~\bibnamefont {Clarke}}, \bibinfo {author} {\bibfnamefont
  {M.}~\bibnamefont {Courthold}}, \bibinfo {author} {\bibfnamefont
  {V.}~\bibnamefont {Bayliss}}, \bibinfo {author} {\bibfnamefont
  {A.}~\bibnamefont {Bernhard}}, \bibinfo {author} {\bibfnamefont
  {M.}~\bibnamefont {Trunk}}, \bibinfo {author} {\bibfnamefont
  {M.}~\bibnamefont {Valléau}}, \bibinfo {author} {\bibfnamefont
  {O.}~\bibnamefont {Marcouillé}}, \bibinfo {author} {\bibfnamefont
  {A.}~\bibnamefont {Chancé}}, \bibinfo {author} {\bibfnamefont
  {S.}~\bibnamefont {Licciardi}}, \bibinfo {author} {\bibfnamefont
  {V.}~\bibnamefont {Malka}}, \bibinfo {author} {\bibfnamefont
  {F.}~\bibnamefont {Nguyen}}, \ and\ \bibinfo {author} {\bibfnamefont
  {G.}~\bibnamefont {Dattoli}},\ }\href {\doibase
  https://doi.org/10.1016/j.physrep.2021.09.001} {\bibfield  {journal}
  {\bibinfo  {journal} {Physics Reports}\ }\textbf {\bibinfo {volume} {937}},\
  \bibinfo {pages} {1} (\bibinfo {year} {2021})}\BibitemShut {NoStop}%
\bibitem [{\citenamefont {Duris}\ \emph {et~al.}(2020)\citenamefont {Duris},
  \citenamefont {Li}, \citenamefont {Driver}, \citenamefont {Champenois},
  \citenamefont {MacArthur}, \citenamefont {Lutman}, \citenamefont {Zhang},
  \citenamefont {Rosenberger}, \citenamefont {Aldrich}, \citenamefont {Coffee},
  \citenamefont {Coslovich}, \citenamefont {Decker}, \citenamefont {Glownia},
  \citenamefont {Hartmann}, \citenamefont {Helml}, \citenamefont {Kamalov},
  \citenamefont {Knurr}, \citenamefont {Krzywinski}, \citenamefont {Lin},
  \citenamefont {Marangos}, \citenamefont {Nantel}, \citenamefont {Natan},
  \citenamefont {O’Neal}, \citenamefont {Shivaram}, \citenamefont {Walter},
  \citenamefont {Wang}, \citenamefont {Welch}, \citenamefont {Wolf},
  \citenamefont {Xu}, \citenamefont {Kling}, \citenamefont {Bucksbaum},
  \citenamefont {Zholents}, \citenamefont {Huang}, \citenamefont {Cryan},\ and\
  \citenamefont {Marinelli}}]{undulatorDuris2019TunableIA}%
  \BibitemOpen
  \bibfield  {author} {\bibinfo {author} {\bibfnamefont {J.}~\bibnamefont
  {Duris}}, \bibinfo {author} {\bibfnamefont {S.}~\bibnamefont {Li}}, \bibinfo
  {author} {\bibfnamefont {T.}~\bibnamefont {Driver}}, \bibinfo {author}
  {\bibfnamefont {E.~G.}\ \bibnamefont {Champenois}}, \bibinfo {author}
  {\bibfnamefont {J.~P.}\ \bibnamefont {MacArthur}}, \bibinfo {author}
  {\bibfnamefont {A.~A.}\ \bibnamefont {Lutman}}, \bibinfo {author}
  {\bibfnamefont {Z.}~\bibnamefont {Zhang}}, \bibinfo {author} {\bibfnamefont
  {P.}~\bibnamefont {Rosenberger}}, \bibinfo {author} {\bibfnamefont {J.~W.}\
  \bibnamefont {Aldrich}}, \bibinfo {author} {\bibfnamefont {R.}~\bibnamefont
  {Coffee}}, \bibinfo {author} {\bibfnamefont {G.}~\bibnamefont {Coslovich}},
  \bibinfo {author} {\bibfnamefont {F.-J.}\ \bibnamefont {Decker}}, \bibinfo
  {author} {\bibfnamefont {J.~M.}\ \bibnamefont {Glownia}}, \bibinfo {author}
  {\bibfnamefont {G.}~\bibnamefont {Hartmann}}, \bibinfo {author}
  {\bibfnamefont {W.}~\bibnamefont {Helml}}, \bibinfo {author} {\bibfnamefont
  {A.}~\bibnamefont {Kamalov}}, \bibinfo {author} {\bibfnamefont
  {J.}~\bibnamefont {Knurr}}, \bibinfo {author} {\bibfnamefont
  {J.}~\bibnamefont {Krzywinski}}, \bibinfo {author} {\bibfnamefont {M.-F.}\
  \bibnamefont {Lin}}, \bibinfo {author} {\bibfnamefont {J.~P.}\ \bibnamefont
  {Marangos}}, \bibinfo {author} {\bibfnamefont {M.}~\bibnamefont {Nantel}},
  \bibinfo {author} {\bibfnamefont {A.}~\bibnamefont {Natan}}, \bibinfo
  {author} {\bibfnamefont {J.~T.}\ \bibnamefont {O’Neal}}, \bibinfo {author}
  {\bibfnamefont {N.}~\bibnamefont {Shivaram}}, \bibinfo {author}
  {\bibfnamefont {P.}~\bibnamefont {Walter}}, \bibinfo {author} {\bibfnamefont
  {A.~L.}\ \bibnamefont {Wang}}, \bibinfo {author} {\bibfnamefont {J.~J.}\
  \bibnamefont {Welch}}, \bibinfo {author} {\bibfnamefont {T.~J.~A.}\
  \bibnamefont {Wolf}}, \bibinfo {author} {\bibfnamefont {J.~Z.}\ \bibnamefont
  {Xu}}, \bibinfo {author} {\bibfnamefont {M.~F.}\ \bibnamefont {Kling}},
  \bibinfo {author} {\bibfnamefont {P.~H.}\ \bibnamefont {Bucksbaum}}, \bibinfo
  {author} {\bibfnamefont {A.}~\bibnamefont {Zholents}}, \bibinfo {author}
  {\bibfnamefont {Z.}~\bibnamefont {Huang}}, \bibinfo {author} {\bibfnamefont
  {J.~P.}\ \bibnamefont {Cryan}}, \ and\ \bibinfo {author} {\bibfnamefont
  {A.}~\bibnamefont {Marinelli}},\ }\href {\doibase 10.1038/s41566-019-0549-5}
  {\bibfield  {journal} {\bibinfo  {journal} {Nature Photonics}\ }\textbf
  {\bibinfo {volume} {14}},\ \bibinfo {pages} {30} (\bibinfo {year}
  {2020})}\BibitemShut {NoStop}%
\bibitem [{\citenamefont {Huang}\ \emph {et~al.}(2017)\citenamefont {Huang},
  \citenamefont {Ding}, \citenamefont {Feng}, \citenamefont {Hemsing},
  \citenamefont {Huang}, \citenamefont {Krzywinski}, \citenamefont {Lutman},
  \citenamefont {Marinelli}, \citenamefont {Maxwell},\ and\ \citenamefont
  {Zhu}}]{undulatorPhysRevLett.119.154801}%
  \BibitemOpen
  \bibfield  {author} {\bibinfo {author} {\bibfnamefont {S.}~\bibnamefont
  {Huang}}, \bibinfo {author} {\bibfnamefont {Y.}~\bibnamefont {Ding}},
  \bibinfo {author} {\bibfnamefont {Y.}~\bibnamefont {Feng}}, \bibinfo {author}
  {\bibfnamefont {E.}~\bibnamefont {Hemsing}}, \bibinfo {author} {\bibfnamefont
  {Z.}~\bibnamefont {Huang}}, \bibinfo {author} {\bibfnamefont
  {J.}~\bibnamefont {Krzywinski}}, \bibinfo {author} {\bibfnamefont
  {A.}~\bibnamefont {Lutman}}, \bibinfo {author} {\bibfnamefont
  {A.}~\bibnamefont {Marinelli}}, \bibinfo {author} {\bibfnamefont
  {T.}~\bibnamefont {Maxwell}}, \ and\ \bibinfo {author} {\bibfnamefont
  {D.}~\bibnamefont {Zhu}},\ }\href {\doibase 10.1103/PhysRevLett.119.154801}
  {\bibfield  {journal} {\bibinfo  {journal} {Phys. Rev. Lett.}\ }\textbf
  {\bibinfo {volume} {119}},\ \bibinfo {pages} {154801} (\bibinfo {year}
  {2017})}\BibitemShut {NoStop}%
\bibitem [{\citenamefont {Xu}\ \emph {et~al.}(2023)\citenamefont {Xu},
  \citenamefont {Liu}, \citenamefont {Dalichaouch}, \citenamefont {Tsung},
  \citenamefont {Zhang}, \citenamefont {Huang}, \citenamefont {Hogan},
  \citenamefont {Yan}, \citenamefont {Joshi},\ and\ \citenamefont
  {Mori}}]{xuUltracompactAttosecondXray2023}%
  \BibitemOpen
  \bibfield  {author} {\bibinfo {author} {\bibfnamefont {X.}~\bibnamefont
  {Xu}}, \bibinfo {author} {\bibfnamefont {J.}~\bibnamefont {Liu}}, \bibinfo
  {author} {\bibfnamefont {T.}~\bibnamefont {Dalichaouch}}, \bibinfo {author}
  {\bibfnamefont {F.~S.}\ \bibnamefont {Tsung}}, \bibinfo {author}
  {\bibfnamefont {Z.}~\bibnamefont {Zhang}}, \bibinfo {author} {\bibfnamefont
  {Z.}~\bibnamefont {Huang}}, \bibinfo {author} {\bibfnamefont {M.~J.}\
  \bibnamefont {Hogan}}, \bibinfo {author} {\bibfnamefont {X.}~\bibnamefont
  {Yan}}, \bibinfo {author} {\bibfnamefont {C.}~\bibnamefont {Joshi}}, \ and\
  \bibinfo {author} {\bibfnamefont {W.~B.}\ \bibnamefont {Mori}},\ }\href
  {\doibase 10.48550/arXiv.2302.08864} {\enquote {\bibinfo {title}
  {Ultra-compact attosecond {X}-ray free-electron lasers utilizing unique beams
  from plasma-based acceleration and an optical undulator},}\ } (\bibinfo
  {year} {2023}),\ \bibinfo {note} {arXiv:2302.08864
  [physics.plasm-ph]}\BibitemShut {NoStop}%
\bibitem [{\citenamefont {Morimoto}\ and\ \citenamefont
  {Baum}(2018)}]{morimoto_diffraction_2018}%
  \BibitemOpen
  \bibfield  {author} {\bibinfo {author} {\bibfnamefont {Y.}~\bibnamefont
  {Morimoto}}\ and\ \bibinfo {author} {\bibfnamefont {P.}~\bibnamefont
  {Baum}},\ }\href {\doibase 10.1038/s41567-017-0007-6} {\bibfield  {journal}
  {\bibinfo  {journal} {Nature Physics}\ }\textbf {\bibinfo {volume} {14}},\
  \bibinfo {pages} {252} (\bibinfo {year} {2018})}\BibitemShut {NoStop}%
\bibitem [{\citenamefont {Nabben}\ \emph {et~al.}(2023)\citenamefont {Nabben},
  \citenamefont {Kuttruff}, \citenamefont {Stolz}, \citenamefont {Ryabov},\
  and\ \citenamefont {Baum}}]{nabben_attosecond_2023}%
  \BibitemOpen
  \bibfield  {author} {\bibinfo {author} {\bibfnamefont {D.}~\bibnamefont
  {Nabben}}, \bibinfo {author} {\bibfnamefont {J.}~\bibnamefont {Kuttruff}},
  \bibinfo {author} {\bibfnamefont {L.}~\bibnamefont {Stolz}}, \bibinfo
  {author} {\bibfnamefont {A.}~\bibnamefont {Ryabov}}, \ and\ \bibinfo {author}
  {\bibfnamefont {P.}~\bibnamefont {Baum}},\ }\href {\doibase
  10.1038/s41586-023-06074-9} {\bibfield  {journal} {\bibinfo  {journal}
  {Nature}\ }\textbf {\bibinfo {volume} {619}},\ \bibinfo {pages} {63}
  (\bibinfo {year} {2023})}\BibitemShut {NoStop}%
\bibitem [{\citenamefont {Hu}\ \emph {et~al.}(2021)\citenamefont {Hu},
  \citenamefont {Zhao}, \citenamefont {Zhang}, \citenamefont {Lu},
  \citenamefont {Wang}, \citenamefont {Hu}, \citenamefont {Shao},\ and\
  \citenamefont {Yu}}]{hu_attosecond_2021}%
  \BibitemOpen
  \bibfield  {author} {\bibinfo {author} {\bibfnamefont {Y.-T.}\ \bibnamefont
  {Hu}}, \bibinfo {author} {\bibfnamefont {J.}~\bibnamefont {Zhao}}, \bibinfo
  {author} {\bibfnamefont {H.}~\bibnamefont {Zhang}}, \bibinfo {author}
  {\bibfnamefont {Y.}~\bibnamefont {Lu}}, \bibinfo {author} {\bibfnamefont
  {W.-Q.}\ \bibnamefont {Wang}}, \bibinfo {author} {\bibfnamefont {L.-X.}\
  \bibnamefont {Hu}}, \bibinfo {author} {\bibfnamefont {F.-Q.}\ \bibnamefont
  {Shao}}, \ and\ \bibinfo {author} {\bibfnamefont {T.-P.}\ \bibnamefont
  {Yu}},\ }\href {\doibase 10.1063/5.0028203} {\bibfield  {journal} {\bibinfo
  {journal} {Applied Physics Letters}\ }\textbf {\bibinfo {volume} {118}},\
  \bibinfo {pages} {054101} (\bibinfo {year} {2021})}\BibitemShut {NoStop}%
\bibitem [{\citenamefont {Zhao}\ \emph {et~al.}(2022)\citenamefont {Zhao},
  \citenamefont {Hu}, \citenamefont {Lu}, \citenamefont {Zhang}, \citenamefont
  {Hu}, \citenamefont {Zhu}, \citenamefont {Sheng}, \citenamefont {Turcu},
  \citenamefont {Pukhov}, \citenamefont {Shao},\ and\ \citenamefont
  {Yu}}]{zhao_all-optical_2022}%
  \BibitemOpen
  \bibfield  {author} {\bibinfo {author} {\bibfnamefont {J.}~\bibnamefont
  {Zhao}}, \bibinfo {author} {\bibfnamefont {Y.-T.}\ \bibnamefont {Hu}},
  \bibinfo {author} {\bibfnamefont {Y.}~\bibnamefont {Lu}}, \bibinfo {author}
  {\bibfnamefont {H.}~\bibnamefont {Zhang}}, \bibinfo {author} {\bibfnamefont
  {L.-X.}\ \bibnamefont {Hu}}, \bibinfo {author} {\bibfnamefont {X.-L.}\
  \bibnamefont {Zhu}}, \bibinfo {author} {\bibfnamefont {Z.-M.}\ \bibnamefont
  {Sheng}}, \bibinfo {author} {\bibfnamefont {I.~C.~E.}\ \bibnamefont {Turcu}},
  \bibinfo {author} {\bibfnamefont {A.}~\bibnamefont {Pukhov}}, \bibinfo
  {author} {\bibfnamefont {F.-Q.}\ \bibnamefont {Shao}}, \ and\ \bibinfo
  {author} {\bibfnamefont {T.-P.}\ \bibnamefont {Yu}},\ }\href {\doibase
  10.1038/s42005-021-00797-9} {\bibfield  {journal} {\bibinfo  {journal}
  {Communications Physics}\ }\textbf {\bibinfo {volume} {5}},\ \bibinfo {pages}
  {15} (\bibinfo {year} {2022})}\BibitemShut {NoStop}%
\bibitem [{\citenamefont {Li}\ \emph {et~al.}(2017)\citenamefont {Li},
  \citenamefont {Yu}, \citenamefont {Hu}, \citenamefont {Yin}, \citenamefont
  {Zou}, \citenamefont {Liu}, \citenamefont {Wang}, \citenamefont {Hu},\ and\
  \citenamefont {Shao}}]{li_ultra-bright_2017}%
  \BibitemOpen
  \bibfield  {author} {\bibinfo {author} {\bibfnamefont {H.-Z.}\ \bibnamefont
  {Li}}, \bibinfo {author} {\bibfnamefont {T.-P.}\ \bibnamefont {Yu}}, \bibinfo
  {author} {\bibfnamefont {L.-X.}\ \bibnamefont {Hu}}, \bibinfo {author}
  {\bibfnamefont {Y.}~\bibnamefont {Yin}}, \bibinfo {author} {\bibfnamefont
  {D.-B.}\ \bibnamefont {Zou}}, \bibinfo {author} {\bibfnamefont {J.-X.}\
  \bibnamefont {Liu}}, \bibinfo {author} {\bibfnamefont {W.-Q.}\ \bibnamefont
  {Wang}}, \bibinfo {author} {\bibfnamefont {S.}~\bibnamefont {Hu}}, \ and\
  \bibinfo {author} {\bibfnamefont {F.-Q.}\ \bibnamefont {Shao}},\ }\href
  {\doibase 10.1364/OE.25.021583} {\bibfield  {journal} {\bibinfo  {journal}
  {Optics Express}\ }\textbf {\bibinfo {volume} {25}},\ \bibinfo {pages}
  {21583} (\bibinfo {year} {2017})}\BibitemShut {NoStop}%
\bibitem [{\citenamefont {Zhu}\ \emph {et~al.}(2019)\citenamefont {Zhu},
  \citenamefont {Chen}, \citenamefont {Yu}, \citenamefont {Weng}, \citenamefont
  {He},\ and\ \citenamefont {Sheng}}]{zhuCollimatedGeVAttosecond2019}%
  \BibitemOpen
  \bibfield  {author} {\bibinfo {author} {\bibfnamefont {X.-L.}\ \bibnamefont
  {Zhu}}, \bibinfo {author} {\bibfnamefont {M.}~\bibnamefont {Chen}}, \bibinfo
  {author} {\bibfnamefont {T.-P.}\ \bibnamefont {Yu}}, \bibinfo {author}
  {\bibfnamefont {S.-M.}\ \bibnamefont {Weng}}, \bibinfo {author}
  {\bibfnamefont {F.}~\bibnamefont {He}}, \ and\ \bibinfo {author}
  {\bibfnamefont {Z.-M.}\ \bibnamefont {Sheng}},\ }\href {\doibase
  10.1063/1.5083914} {\bibfield  {journal} {\bibinfo  {journal} {Matter and
  Radiation at Extremes}\ }\textbf {\bibinfo {volume} {4}},\ \bibinfo {pages}
  {014401} (\bibinfo {year} {2019})}\BibitemShut {NoStop}%
\bibitem [{\citenamefont {Zhang}\ \emph {et~al.}(2022)\citenamefont {Zhang},
  \citenamefont {Liu}, \citenamefont {Tang}, \citenamefont {Luo}, \citenamefont
  {Zhao}, \citenamefont {Zhang},\ and\ \citenamefont
  {Yu}}]{zhang_generation_2022}%
  \BibitemOpen
  \bibfield  {author} {\bibinfo {author} {\bibfnamefont {L.-Q.}\ \bibnamefont
  {Zhang}}, \bibinfo {author} {\bibfnamefont {K.}~\bibnamefont {Liu}}, \bibinfo
  {author} {\bibfnamefont {S.}~\bibnamefont {Tang}}, \bibinfo {author}
  {\bibfnamefont {W.}~\bibnamefont {Luo}}, \bibinfo {author} {\bibfnamefont
  {J.}~\bibnamefont {Zhao}}, \bibinfo {author} {\bibfnamefont {H.}~\bibnamefont
  {Zhang}}, \ and\ \bibinfo {author} {\bibfnamefont {T.-P.}\ \bibnamefont
  {Yu}},\ }\href {\doibase 10.1088/1361-6587/ac85a7} {\bibfield  {journal}
  {\bibinfo  {journal} {Plasma Physics and Controlled Fusion}\ }\textbf
  {\bibinfo {volume} {64}},\ \bibinfo {pages} {105011} (\bibinfo {year}
  {2022})}\BibitemShut {NoStop}%
\bibitem [{\citenamefont {Hu}\ \emph {et~al.}(2024)\citenamefont {Hu},
  \citenamefont {Yu}, \citenamefont {Cao}, \citenamefont {Chen}, \citenamefont
  {Zou}, \citenamefont {Yin}, \citenamefont {Sheng},\ and\ \citenamefont
  {Shao}}]{huRotatingAttosecondElectron2024}%
  \BibitemOpen
  \bibfield  {author} {\bibinfo {author} {\bibfnamefont {L.-X.}\ \bibnamefont
  {Hu}}, \bibinfo {author} {\bibfnamefont {T.-P.}\ \bibnamefont {Yu}}, \bibinfo
  {author} {\bibfnamefont {Y.}~\bibnamefont {Cao}}, \bibinfo {author}
  {\bibfnamefont {M.}~\bibnamefont {Chen}}, \bibinfo {author} {\bibfnamefont
  {D.-B.}\ \bibnamefont {Zou}}, \bibinfo {author} {\bibfnamefont
  {Y.}~\bibnamefont {Yin}}, \bibinfo {author} {\bibfnamefont {Z.-M.}\
  \bibnamefont {Sheng}}, \ and\ \bibinfo {author} {\bibfnamefont {F.-Q.}\
  \bibnamefont {Shao}},\ }\href {\doibase 10.1017/hpl.2024.66} {\bibfield
  {journal} {\bibinfo  {journal} {High Power Laser Science and Engineering}\
  }\textbf {\bibinfo {volume} {12}},\ \bibinfo {pages} {e69} (\bibinfo {year}
  {2024})}\BibitemShut {NoStop}%
\bibitem [{\citenamefont {Huang}\ \emph {et~al.}(2021)\citenamefont {Huang},
  \citenamefont {Deng}, \citenamefont {Liu}, \citenamefont {Wang},\ and\
  \citenamefont {Zhao}}]{huang_features_2021}%
  \BibitemOpen
  \bibfield  {author} {\bibinfo {author} {\bibfnamefont {N.}~\bibnamefont
  {Huang}}, \bibinfo {author} {\bibfnamefont {H.}~\bibnamefont {Deng}},
  \bibinfo {author} {\bibfnamefont {B.}~\bibnamefont {Liu}}, \bibinfo {author}
  {\bibfnamefont {D.}~\bibnamefont {Wang}}, \ and\ \bibinfo {author}
  {\bibfnamefont {Z.}~\bibnamefont {Zhao}},\ }\href {\doibase
  10.1016/j.xinn.2021.100097} {\bibfield  {journal} {\bibinfo  {journal} {The
  Innovation}\ }\textbf {\bibinfo {volume} {2}},\ \bibinfo {pages} {100097}
  (\bibinfo {year} {2021})}\BibitemShut {NoStop}%
\bibitem [{\citenamefont {Petrillo}\ \emph {et~al.}(2008)\citenamefont
  {Petrillo}, \citenamefont {Serafini},\ and\ \citenamefont
  {Tomassini}}]{petrillo_ultrahigh_2008}%
  \BibitemOpen
  \bibfield  {author} {\bibinfo {author} {\bibfnamefont {V.}~\bibnamefont
  {Petrillo}}, \bibinfo {author} {\bibfnamefont {L.}~\bibnamefont {Serafini}},
  \ and\ \bibinfo {author} {\bibfnamefont {P.}~\bibnamefont {Tomassini}},\
  }\href {\doibase 10.1103/PhysRevSTAB.11.070703} {\bibfield  {journal}
  {\bibinfo  {journal} {Physical Review Special Topics - Accelerators and
  Beams}\ }\textbf {\bibinfo {volume} {11}},\ \bibinfo {pages} {070703}
  (\bibinfo {year} {2008})}\BibitemShut {NoStop}%
\bibitem [{\citenamefont {Xu}\ \emph {et~al.}(2022)\citenamefont {Xu},
  \citenamefont {Li}, \citenamefont {Tsung}, \citenamefont {Miller},
  \citenamefont {Yakimenko}, \citenamefont {Hogan}, \citenamefont {Joshi},\
  and\ \citenamefont {Mori}}]{xu_generation_2022}%
  \BibitemOpen
  \bibfield  {author} {\bibinfo {author} {\bibfnamefont {X.}~\bibnamefont
  {Xu}}, \bibinfo {author} {\bibfnamefont {F.}~\bibnamefont {Li}}, \bibinfo
  {author} {\bibfnamefont {F.~S.}\ \bibnamefont {Tsung}}, \bibinfo {author}
  {\bibfnamefont {K.}~\bibnamefont {Miller}}, \bibinfo {author} {\bibfnamefont
  {V.}~\bibnamefont {Yakimenko}}, \bibinfo {author} {\bibfnamefont {M.~J.}\
  \bibnamefont {Hogan}}, \bibinfo {author} {\bibfnamefont {C.}~\bibnamefont
  {Joshi}}, \ and\ \bibinfo {author} {\bibfnamefont {W.~B.}\ \bibnamefont
  {Mori}},\ }\href {\doibase 10.1038/s41467-022-30806-6} {\bibfield  {journal}
  {\bibinfo  {journal} {Nature Communications}\ }\textbf {\bibinfo {volume}
  {13}},\ \bibinfo {pages} {3364} (\bibinfo {year} {2022})}\BibitemShut
  {NoStop}%
\bibitem [{\citenamefont {Feng}\ \emph {et~al.}(2024)\citenamefont {Feng},
  \citenamefont {Jiang}, \citenamefont {Hu}, \citenamefont {Luan},
  \citenamefont {Wang},\ and\ \citenamefont
  {Li}}]{fengBunchingEnhancementCoherent2024}%
  \BibitemOpen
  \bibfield  {author} {\bibinfo {author} {\bibfnamefont {K.}~\bibnamefont
  {Feng}}, \bibinfo {author} {\bibfnamefont {K.}~\bibnamefont {Jiang}},
  \bibinfo {author} {\bibfnamefont {R.}~\bibnamefont {Hu}}, \bibinfo {author}
  {\bibfnamefont {S.}~\bibnamefont {Luan}}, \bibinfo {author} {\bibfnamefont
  {W.}~\bibnamefont {Wang}}, \ and\ \bibinfo {author} {\bibfnamefont
  {R.}~\bibnamefont {Li}},\ }\href {\doibase 10.1063/5.0191508} {\bibfield
  {journal} {\bibinfo  {journal} {Matter and Radiation at Extremes}\ }\textbf
  {\bibinfo {volume} {9}},\ \bibinfo {pages} {057201} (\bibinfo {year}
  {2024})}\BibitemShut {NoStop}%
\bibitem [{\citenamefont {Zhang}\ \emph {et~al.}(2024)\citenamefont {Zhang},
  \citenamefont {Hu}, \citenamefont {Cao}, \citenamefont {Shao},\ and\
  \citenamefont {Yu}}]{zhang_generation_2024}%
  \BibitemOpen
  \bibfield  {author} {\bibinfo {author} {\bibfnamefont {W.~Y.}\ \bibnamefont
  {Zhang}}, \bibinfo {author} {\bibfnamefont {L.~X.}\ \bibnamefont {Hu}},
  \bibinfo {author} {\bibfnamefont {Y.}~\bibnamefont {Cao}}, \bibinfo {author}
  {\bibfnamefont {F.~Q.}\ \bibnamefont {Shao}}, \ and\ \bibinfo {author}
  {\bibfnamefont {T.~P.}\ \bibnamefont {Yu}},\ }\href {\doibase
  10.1364/OE.521360} {\bibfield  {journal} {\bibinfo  {journal} {Optics
  Express}\ }\textbf {\bibinfo {volume} {32}},\ \bibinfo {pages} {16398}
  (\bibinfo {year} {2024})}\BibitemShut {NoStop}%
\bibitem [{\citenamefont {Yu}\ \emph {et~al.}(2024)\citenamefont {Yu},
  \citenamefont {Liu}, \citenamefont {Zhao}, \citenamefont {Zhu}, \citenamefont
  {Lu}, \citenamefont {Cao}, \citenamefont {Zhang}, \citenamefont {Shao},\ and\
  \citenamefont {Sheng}}]{yu_bright_2024}%
  \BibitemOpen
  \bibfield  {author} {\bibinfo {author} {\bibfnamefont {T.-P.}\ \bibnamefont
  {Yu}}, \bibinfo {author} {\bibfnamefont {K.}~\bibnamefont {Liu}}, \bibinfo
  {author} {\bibfnamefont {J.}~\bibnamefont {Zhao}}, \bibinfo {author}
  {\bibfnamefont {X.-L.}\ \bibnamefont {Zhu}}, \bibinfo {author} {\bibfnamefont
  {Y.}~\bibnamefont {Lu}}, \bibinfo {author} {\bibfnamefont {Y.}~\bibnamefont
  {Cao}}, \bibinfo {author} {\bibfnamefont {H.}~\bibnamefont {Zhang}}, \bibinfo
  {author} {\bibfnamefont {F.-Q.}\ \bibnamefont {Shao}}, \ and\ \bibinfo
  {author} {\bibfnamefont {Z.-M.}\ \bibnamefont {Sheng}},\ }\href {\doibase
  10.1007/s41614-024-00158-3} {\bibfield  {journal} {\bibinfo  {journal}
  {Reviews of Modern Plasma Physics}\ }\textbf {\bibinfo {volume} {8}},\
  \bibinfo {pages} {24} (\bibinfo {year} {2024})}\BibitemShut {NoStop}%
\bibitem [{\citenamefont {Jiang}\ \emph {et~al.}(2021)\citenamefont {Jiang},
  \citenamefont {Wang}, \citenamefont {Weber}, \citenamefont {Dong},
  \citenamefont {Leng}, \citenamefont {Li},\ and\ \citenamefont
  {Xu}}]{Jiang2021DirectAO}%
  \BibitemOpen
  \bibfield  {author} {\bibinfo {author} {\bibfnamefont {C.}~\bibnamefont
  {Jiang}}, \bibinfo {author} {\bibfnamefont {W.~P.}\ \bibnamefont {Wang}},
  \bibinfo {author} {\bibfnamefont {S.}~\bibnamefont {Weber}}, \bibinfo
  {author} {\bibfnamefont {H.}~\bibnamefont {Dong}}, \bibinfo {author}
  {\bibfnamefont {Y.~X.}\ \bibnamefont {Leng}}, \bibinfo {author}
  {\bibfnamefont {R.~X.}\ \bibnamefont {Li}}, \ and\ \bibinfo {author}
  {\bibfnamefont {Z.~Z.}\ \bibnamefont {Xu}},\ }\href {\doibase
  10.1017/hpl.2021.28} {\bibfield  {journal} {\bibinfo  {journal} {High Power
  Laser Science and Engineering}\ }\textbf {\bibinfo {volume} {9}},\ \bibinfo
  {pages} {e44} (\bibinfo {year} {2021})}\BibitemShut {NoStop}%
\bibitem [{\citenamefont {Liang}\ \emph {et~al.}(2020)\citenamefont {Liang},
  \citenamefont {Shen}, \citenamefont {Zhang},\ and\ \citenamefont
  {Zhang}}]{liangHighrepetitionrateFewattosecondHighquality2020}%
  \BibitemOpen
  \bibfield  {author} {\bibinfo {author} {\bibfnamefont {Z.}~\bibnamefont
  {Liang}}, \bibinfo {author} {\bibfnamefont {B.}~\bibnamefont {Shen}},
  \bibinfo {author} {\bibfnamefont {X.}~\bibnamefont {Zhang}}, \ and\ \bibinfo
  {author} {\bibfnamefont {L.}~\bibnamefont {Zhang}},\ }\href {\doibase
  10.1063/5.0004524} {\bibfield  {journal} {\bibinfo  {journal} {Matter and
  Radiation at Extremes}\ }\textbf {\bibinfo {volume} {5}},\ \bibinfo {pages}
  {054401} (\bibinfo {year} {2020})}\BibitemShut {NoStop}%
\bibitem [{\citenamefont {Li}\ \emph {et~al.}(2013)\citenamefont {Li},
  \citenamefont {Sheng}, \citenamefont {Liu}, \citenamefont {Meyer-ter Vehn},
  \citenamefont {Mori}, \citenamefont {Lu},\ and\ \citenamefont
  {Zhang}}]{li_dense_2013}%
  \BibitemOpen
  \bibfield  {author} {\bibinfo {author} {\bibfnamefont {F.~Y.}\ \bibnamefont
  {Li}}, \bibinfo {author} {\bibfnamefont {Z.~M.}\ \bibnamefont {Sheng}},
  \bibinfo {author} {\bibfnamefont {Y.}~\bibnamefont {Liu}}, \bibinfo {author}
  {\bibfnamefont {J.}~\bibnamefont {Meyer-ter Vehn}}, \bibinfo {author}
  {\bibfnamefont {W.~B.}\ \bibnamefont {Mori}}, \bibinfo {author}
  {\bibfnamefont {W.}~\bibnamefont {Lu}}, \ and\ \bibinfo {author}
  {\bibfnamefont {J.}~\bibnamefont {Zhang}},\ }\href {\doibase
  10.1103/PhysRevLett.110.135002} {\bibfield  {journal} {\bibinfo  {journal}
  {Phys. Rev. Lett.}\ }\textbf {\bibinfo {volume} {110}},\ \bibinfo {pages}
  {135002} (\bibinfo {year} {2013})}\BibitemShut {NoStop}%
\bibitem [{\citenamefont {Zhu}\ \emph {et~al.}(2021)\citenamefont {Zhu},
  \citenamefont {Liu}, \citenamefont {Chen}, \citenamefont {Weng},
  \citenamefont {He}, \citenamefont {Assmann}, \citenamefont {Sheng},\ and\
  \citenamefont {Zhang}}]{zhu_generation_2021}%
  \BibitemOpen
  \bibfield  {author} {\bibinfo {author} {\bibfnamefont {X.-L.}\ \bibnamefont
  {Zhu}}, \bibinfo {author} {\bibfnamefont {W.-Y.}\ \bibnamefont {Liu}},
  \bibinfo {author} {\bibfnamefont {M.}~\bibnamefont {Chen}}, \bibinfo {author}
  {\bibfnamefont {S.-M.}\ \bibnamefont {Weng}}, \bibinfo {author}
  {\bibfnamefont {F.}~\bibnamefont {He}}, \bibinfo {author} {\bibfnamefont
  {R.}~\bibnamefont {Assmann}}, \bibinfo {author} {\bibfnamefont {Z.-M.}\
  \bibnamefont {Sheng}}, \ and\ \bibinfo {author} {\bibfnamefont
  {J.}~\bibnamefont {Zhang}},\ }\href {\doibase
  10.1103/PhysRevApplied.15.044039} {\bibfield  {journal} {\bibinfo  {journal}
  {Physical Review Applied}\ }\textbf {\bibinfo {volume} {15}},\ \bibinfo
  {pages} {044039} (\bibinfo {year} {2021})}\BibitemShut {NoStop}%
\bibitem [{\citenamefont {Sun}\ \emph {et~al.}(2024)\citenamefont {Sun},
  \citenamefont {Zhao}, \citenamefont {Wan}, \citenamefont {Salamin},\ and\
  \citenamefont {Li}}]{sun_generation_2024}%
  \BibitemOpen
  \bibfield  {author} {\bibinfo {author} {\bibfnamefont {T.}~\bibnamefont
  {Sun}}, \bibinfo {author} {\bibfnamefont {Q.}~\bibnamefont {Zhao}}, \bibinfo
  {author} {\bibfnamefont {F.}~\bibnamefont {Wan}}, \bibinfo {author}
  {\bibfnamefont {Y.~I.}\ \bibnamefont {Salamin}}, \ and\ \bibinfo {author}
  {\bibfnamefont {J.-X.}\ \bibnamefont {Li}},\ }\href {\doibase
  10.1103/PhysRevLett.132.045001} {\bibfield  {journal} {\bibinfo  {journal}
  {Physical Review Letters}\ }\textbf {\bibinfo {volume} {132}},\ \bibinfo
  {pages} {045001} (\bibinfo {year} {2024})}\BibitemShut {NoStop}%
\bibitem [{\citenamefont {Deng}\ \emph {et~al.}(2023)\citenamefont {Deng},
  \citenamefont {Li}, \citenamefont {Luo}, \citenamefont {Li},\ and\
  \citenamefont {Zeng}}]{deng_generation_2023}%
  \BibitemOpen
  \bibfield  {author} {\bibinfo {author} {\bibfnamefont {A.}~\bibnamefont
  {Deng}}, \bibinfo {author} {\bibfnamefont {X.}~\bibnamefont {Li}}, \bibinfo
  {author} {\bibfnamefont {Z.}~\bibnamefont {Luo}}, \bibinfo {author}
  {\bibfnamefont {Y.}~\bibnamefont {Li}}, \ and\ \bibinfo {author}
  {\bibfnamefont {J.}~\bibnamefont {Zeng}},\ }\href {\doibase
  10.1364/OE.492468} {\bibfield  {journal} {\bibinfo  {journal} {Optics
  Express}\ }\textbf {\bibinfo {volume} {31}},\ \bibinfo {pages} {19958}
  (\bibinfo {year} {2023})}\BibitemShut {NoStop}%
\bibitem [{\citenamefont {Weikum}\ \emph {et~al.}(2016)\citenamefont {Weikum},
  \citenamefont {Li}, \citenamefont {Assmann}, \citenamefont {Sheng},\ and\
  \citenamefont {Jaroszynski}}]{weikum_generation_2016}%
  \BibitemOpen
  \bibfield  {author} {\bibinfo {author} {\bibfnamefont {M.}~\bibnamefont
  {Weikum}}, \bibinfo {author} {\bibfnamefont {F.}~\bibnamefont {Li}}, \bibinfo
  {author} {\bibfnamefont {R.}~\bibnamefont {Assmann}}, \bibinfo {author}
  {\bibfnamefont {Z.}~\bibnamefont {Sheng}}, \ and\ \bibinfo {author}
  {\bibfnamefont {D.}~\bibnamefont {Jaroszynski}},\ }\href {\doibase
  10.1016/j.nima.2016.01.003} {\bibfield  {journal} {\bibinfo  {journal}
  {Nuclear Instruments and Methods in Physics Research Section A: Accelerators,
  Spectrometers, Detectors and Associated Equipment}\ }\textbf {\bibinfo
  {volume} {829}},\ \bibinfo {pages} {33} (\bibinfo {year} {2016})}\BibitemShut
  {NoStop}%
\bibitem [{\citenamefont {Dodin}\ and\ \citenamefont
  {Fisch}(2007)}]{dodinStochasticExtractionPeriodic2007}%
  \BibitemOpen
  \bibfield  {author} {\bibinfo {author} {\bibfnamefont {I.~Y.}\ \bibnamefont
  {Dodin}}\ and\ \bibinfo {author} {\bibfnamefont {N.~J.}\ \bibnamefont
  {Fisch}},\ }\href {\doibase 10.1103/PhysRevLett.98.234801} {\bibfield
  {journal} {\bibinfo  {journal} {Physical Review Letters}\ }\textbf {\bibinfo
  {volume} {98}},\ \bibinfo {pages} {234801} (\bibinfo {year}
  {2007})}\BibitemShut {NoStop}%
\bibitem [{\citenamefont {Hu}\ \emph {et~al.}(2018{\natexlab{a}})\citenamefont
  {Hu}, \citenamefont {Yu}, \citenamefont {Li}, \citenamefont {Yin},
  \citenamefont {McKenna},\ and\ \citenamefont {Shao}}]{hu_dense_2018}%
  \BibitemOpen
  \bibfield  {author} {\bibinfo {author} {\bibfnamefont {L.-X.}\ \bibnamefont
  {Hu}}, \bibinfo {author} {\bibfnamefont {T.-P.}\ \bibnamefont {Yu}}, \bibinfo
  {author} {\bibfnamefont {H.-Z.}\ \bibnamefont {Li}}, \bibinfo {author}
  {\bibfnamefont {Y.}~\bibnamefont {Yin}}, \bibinfo {author} {\bibfnamefont
  {P.}~\bibnamefont {McKenna}}, \ and\ \bibinfo {author} {\bibfnamefont
  {F.-Q.}\ \bibnamefont {Shao}},\ }\href {\doibase 10.1364/OL.43.002615}
  {\bibfield  {journal} {\bibinfo  {journal} {Optics Letters}\ }\textbf
  {\bibinfo {volume} {43}},\ \bibinfo {pages} {2615} (\bibinfo {year}
  {2018}{\natexlab{a}})}\BibitemShut {NoStop}%
\bibitem [{\citenamefont {Hu}\ \emph {et~al.}(2018{\natexlab{b}})\citenamefont
  {Hu}, \citenamefont {Yu}, \citenamefont {Sheng}, \citenamefont {Vieira},
  \citenamefont {Zou}, \citenamefont {Yin}, \citenamefont {McKenna},\ and\
  \citenamefont {Shao}}]{hu_attosecond_2018}%
  \BibitemOpen
  \bibfield  {author} {\bibinfo {author} {\bibfnamefont {L.-X.}\ \bibnamefont
  {Hu}}, \bibinfo {author} {\bibfnamefont {T.-P.}\ \bibnamefont {Yu}}, \bibinfo
  {author} {\bibfnamefont {Z.-M.}\ \bibnamefont {Sheng}}, \bibinfo {author}
  {\bibfnamefont {J.}~\bibnamefont {Vieira}}, \bibinfo {author} {\bibfnamefont
  {D.-B.}\ \bibnamefont {Zou}}, \bibinfo {author} {\bibfnamefont
  {Y.}~\bibnamefont {Yin}}, \bibinfo {author} {\bibfnamefont {P.}~\bibnamefont
  {McKenna}}, \ and\ \bibinfo {author} {\bibfnamefont {F.-Q.}\ \bibnamefont
  {Shao}},\ }\href {\doibase 10.1038/s41598-018-25421-9} {\bibfield  {journal}
  {\bibinfo  {journal} {Scientific Reports}\ }\textbf {\bibinfo {volume} {8}},\
  \bibinfo {pages} {7282} (\bibinfo {year} {2018}{\natexlab{b}})}\BibitemShut
  {NoStop}%
\bibitem [{\citenamefont {Naumova}(2006)}]{naumova_efficient_2006}%
  \BibitemOpen
  \bibfield  {author} {\bibinfo {author} {\bibfnamefont {N.}~\bibnamefont
  {Naumova}},\ }in\ \href {\doibase 10.1063/1.2195198} {\emph {\bibinfo
  {booktitle} {{AIP} {Conference} {Proceedings}}}},\ Vol.\ \bibinfo {volume}
  {827}\ (\bibinfo  {publisher} {AIP},\ \bibinfo {address} {Varenna (Italy)},\
  \bibinfo {year} {2006})\ pp.\ \bibinfo {pages} {65--73},\ \bibinfo {note}
  {iSSN: 0094243X}\BibitemShut {NoStop}%
\bibitem [{\citenamefont {Naumova}\ \emph {et~al.}(2004)\citenamefont
  {Naumova}, \citenamefont {Sokolov}, \citenamefont {Nees}, \citenamefont
  {Maksimchuk}, \citenamefont {Yanovsky},\ and\ \citenamefont
  {Mourou}}]{naumovaAttosecondElectronBunches2004}%
  \BibitemOpen
  \bibfield  {author} {\bibinfo {author} {\bibfnamefont {N.}~\bibnamefont
  {Naumova}}, \bibinfo {author} {\bibfnamefont {I.}~\bibnamefont {Sokolov}},
  \bibinfo {author} {\bibfnamefont {J.}~\bibnamefont {Nees}}, \bibinfo {author}
  {\bibfnamefont {A.}~\bibnamefont {Maksimchuk}}, \bibinfo {author}
  {\bibfnamefont {V.}~\bibnamefont {Yanovsky}}, \ and\ \bibinfo {author}
  {\bibfnamefont {G.}~\bibnamefont {Mourou}},\ }\href {\doibase
  10.1103/PhysRevLett.93.195003} {\bibfield  {journal} {\bibinfo  {journal}
  {Physical Review Letters}\ }\textbf {\bibinfo {volume} {93}},\ \bibinfo
  {pages} {195003} (\bibinfo {year} {2004})}\BibitemShut {NoStop}%
\bibitem [{\citenamefont {Kozák}(2019)}]{kozakAllOpticalSchemeGeneration2019}%
  \BibitemOpen
  \bibfield  {author} {\bibinfo {author} {\bibfnamefont {M.}~\bibnamefont
  {Kozák}},\ }\href {\doibase 10.1103/PhysRevLett.123.203202} {\bibfield
  {journal} {\bibinfo  {journal} {Physical Review Letters}\ }\textbf {\bibinfo
  {volume} {123}},\ \bibinfo {pages} {203202} (\bibinfo {year}
  {2019})}\BibitemShut {NoStop}%
\bibitem [{\citenamefont {Liseykina}\ \emph {et~al.}(2010)\citenamefont
  {Liseykina}, \citenamefont {Pirner},\ and\ \citenamefont
  {Bauer}}]{liseykinaRelativisticAttosecondElectron2010}%
  \BibitemOpen
  \bibfield  {author} {\bibinfo {author} {\bibfnamefont {T.~V.}\ \bibnamefont
  {Liseykina}}, \bibinfo {author} {\bibfnamefont {S.}~\bibnamefont {Pirner}}, \
  and\ \bibinfo {author} {\bibfnamefont {D.}~\bibnamefont {Bauer}},\ }\href
  {\doibase 10.1103/PhysRevLett.104.095002} {\bibfield  {journal} {\bibinfo
  {journal} {Physical Review Letters}\ }\textbf {\bibinfo {volume} {104}},\
  \bibinfo {pages} {095002} (\bibinfo {year} {2010})}\BibitemShut {NoStop}%
\bibitem [{\citenamefont {Luttikhof}\ \emph {et~al.}(2010)\citenamefont
  {Luttikhof}, \citenamefont {Khachatryan}, \citenamefont {Van~Goor},\ and\
  \citenamefont {Boller}}]{luttikhofGeneratingUltrarelativisticAttosecond2010}%
  \BibitemOpen
  \bibfield  {author} {\bibinfo {author} {\bibfnamefont {M.~J.~H.}\
  \bibnamefont {Luttikhof}}, \bibinfo {author} {\bibfnamefont {A.~G.}\
  \bibnamefont {Khachatryan}}, \bibinfo {author} {\bibfnamefont {F.~A.}\
  \bibnamefont {Van~Goor}}, \ and\ \bibinfo {author} {\bibfnamefont {K.-J.}\
  \bibnamefont {Boller}},\ }\href {\doibase 10.1103/PhysRevLett.105.124801}
  {\bibfield  {journal} {\bibinfo  {journal} {Physical Review Letters}\
  }\textbf {\bibinfo {volume} {105}},\ \bibinfo {pages} {124801} (\bibinfo
  {year} {2010})}\BibitemShut {NoStop}%
\bibitem [{\citenamefont {Chen}\ \emph {et~al.}(1985)\citenamefont {Chen},
  \citenamefont {Dawson}, \citenamefont {Huff},\ and\ \citenamefont
  {Katsouleas}}]{chen_acceleration_1985}%
  \BibitemOpen
  \bibfield  {author} {\bibinfo {author} {\bibfnamefont {P.}~\bibnamefont
  {Chen}}, \bibinfo {author} {\bibfnamefont {J.~M.}\ \bibnamefont {Dawson}},
  \bibinfo {author} {\bibfnamefont {R.~W.}\ \bibnamefont {Huff}}, \ and\
  \bibinfo {author} {\bibfnamefont {T.}~\bibnamefont {Katsouleas}},\ }\href
  {\doibase 10.1103/PhysRevLett.54.693} {\bibfield  {journal} {\bibinfo
  {journal} {Physical Review Letters}\ }\textbf {\bibinfo {volume} {54}},\
  \bibinfo {pages} {693} (\bibinfo {year} {1985})}\BibitemShut {NoStop}%
\bibitem [{\citenamefont {Katsouleas}(1986)}]{katsouleas_physical_1986}%
  \BibitemOpen
  \bibfield  {author} {\bibinfo {author} {\bibfnamefont {T.}~\bibnamefont
  {Katsouleas}},\ }\href {\doibase 10.1103/PhysRevA.33.2056} {\bibfield
  {journal} {\bibinfo  {journal} {Physical Review A}\ }\textbf {\bibinfo
  {volume} {33}},\ \bibinfo {pages} {2056} (\bibinfo {year}
  {1986})}\BibitemShut {NoStop}%
\bibitem [{\citenamefont {Lu}\ \emph {et~al.}(2007)\citenamefont {Lu},
  \citenamefont {Tzoufras}, \citenamefont {Joshi}, \citenamefont {Tsung},
  \citenamefont {Mori}, \citenamefont {Vieira}, \citenamefont {Fonseca},\ and\
  \citenamefont {Silva}}]{lu_generating_2007}%
  \BibitemOpen
  \bibfield  {author} {\bibinfo {author} {\bibfnamefont {W.}~\bibnamefont
  {Lu}}, \bibinfo {author} {\bibfnamefont {M.}~\bibnamefont {Tzoufras}},
  \bibinfo {author} {\bibfnamefont {C.}~\bibnamefont {Joshi}}, \bibinfo
  {author} {\bibfnamefont {F.~S.}\ \bibnamefont {Tsung}}, \bibinfo {author}
  {\bibfnamefont {W.~B.}\ \bibnamefont {Mori}}, \bibinfo {author}
  {\bibfnamefont {J.}~\bibnamefont {Vieira}}, \bibinfo {author} {\bibfnamefont
  {R.~A.}\ \bibnamefont {Fonseca}}, \ and\ \bibinfo {author} {\bibfnamefont
  {L.~O.}\ \bibnamefont {Silva}},\ }\href {\doibase
  10.1103/PhysRevSTAB.10.061301} {\bibfield  {journal} {\bibinfo  {journal}
  {Physical Review Special Topics - Accelerators and Beams}\ }\textbf {\bibinfo
  {volume} {10}},\ \bibinfo {pages} {061301} (\bibinfo {year}
  {2007})}\BibitemShut {NoStop}%
\bibitem [{\citenamefont {Litos}\ \emph {et~al.}(2014)\citenamefont {Litos},
  \citenamefont {Adli}, \citenamefont {An}, \citenamefont {Clarke},
  \citenamefont {Clayton}, \citenamefont {Corde}, \citenamefont {Delahaye},
  \citenamefont {England}, \citenamefont {Fisher}, \citenamefont {Frederico},
  \citenamefont {Gessner}, \citenamefont {Green}, \citenamefont {Hogan},
  \citenamefont {Joshi}, \citenamefont {Lu}, \citenamefont {Marsh},
  \citenamefont {Mori}, \citenamefont {Muggli}, \citenamefont
  {Vafaei-Najafabadi}, \citenamefont {Walz}, \citenamefont {White},
  \citenamefont {Wu}, \citenamefont {Yakimenko},\ and\ \citenamefont
  {Yocky}}]{litos_high-efficiency_2014}%
  \BibitemOpen
  \bibfield  {author} {\bibinfo {author} {\bibfnamefont {M.}~\bibnamefont
  {Litos}}, \bibinfo {author} {\bibfnamefont {E.}~\bibnamefont {Adli}},
  \bibinfo {author} {\bibfnamefont {W.}~\bibnamefont {An}}, \bibinfo {author}
  {\bibfnamefont {C.~I.}\ \bibnamefont {Clarke}}, \bibinfo {author}
  {\bibfnamefont {C.~E.}\ \bibnamefont {Clayton}}, \bibinfo {author}
  {\bibfnamefont {S.}~\bibnamefont {Corde}}, \bibinfo {author} {\bibfnamefont
  {J.~P.}\ \bibnamefont {Delahaye}}, \bibinfo {author} {\bibfnamefont {R.~J.}\
  \bibnamefont {England}}, \bibinfo {author} {\bibfnamefont {A.~S.}\
  \bibnamefont {Fisher}}, \bibinfo {author} {\bibfnamefont {J.}~\bibnamefont
  {Frederico}}, \bibinfo {author} {\bibfnamefont {S.}~\bibnamefont {Gessner}},
  \bibinfo {author} {\bibfnamefont {S.~Z.}\ \bibnamefont {Green}}, \bibinfo
  {author} {\bibfnamefont {M.~J.}\ \bibnamefont {Hogan}}, \bibinfo {author}
  {\bibfnamefont {C.}~\bibnamefont {Joshi}}, \bibinfo {author} {\bibfnamefont
  {W.}~\bibnamefont {Lu}}, \bibinfo {author} {\bibfnamefont {K.~A.}\
  \bibnamefont {Marsh}}, \bibinfo {author} {\bibfnamefont {W.~B.}\ \bibnamefont
  {Mori}}, \bibinfo {author} {\bibfnamefont {P.}~\bibnamefont {Muggli}},
  \bibinfo {author} {\bibfnamefont {N.}~\bibnamefont {Vafaei-Najafabadi}},
  \bibinfo {author} {\bibfnamefont {D.}~\bibnamefont {Walz}}, \bibinfo {author}
  {\bibfnamefont {G.}~\bibnamefont {White}}, \bibinfo {author} {\bibfnamefont
  {Z.}~\bibnamefont {Wu}}, \bibinfo {author} {\bibfnamefont {V.}~\bibnamefont
  {Yakimenko}}, \ and\ \bibinfo {author} {\bibfnamefont {G.}~\bibnamefont
  {Yocky}},\ }\href {\doibase 10.1038/nature13882} {\bibfield  {journal}
  {\bibinfo  {journal} {Nature}\ }\textbf {\bibinfo {volume} {515}},\ \bibinfo
  {pages} {92} (\bibinfo {year} {2014})}\BibitemShut {NoStop}%
\bibitem [{\citenamefont {Li}\ \emph {et~al.}(2022)\citenamefont {Li},
  \citenamefont {Dalichaouch}, \citenamefont {Pierce}, \citenamefont {Xu},
  \citenamefont {Tsung}, \citenamefont {Lu}, \citenamefont {Joshi},\ and\
  \citenamefont {Mori}}]{li_ultrabright_2022}%
  \BibitemOpen
  \bibfield  {author} {\bibinfo {author} {\bibfnamefont {F.}~\bibnamefont
  {Li}}, \bibinfo {author} {\bibfnamefont {T.}~\bibnamefont {Dalichaouch}},
  \bibinfo {author} {\bibfnamefont {J.}~\bibnamefont {Pierce}}, \bibinfo
  {author} {\bibfnamefont {X.}~\bibnamefont {Xu}}, \bibinfo {author}
  {\bibfnamefont {F.}~\bibnamefont {Tsung}}, \bibinfo {author} {\bibfnamefont
  {W.}~\bibnamefont {Lu}}, \bibinfo {author} {\bibfnamefont {C.}~\bibnamefont
  {Joshi}}, \ and\ \bibinfo {author} {\bibfnamefont {W.}~\bibnamefont {Mori}},\
  }\href {\doibase 10.1103/PhysRevLett.128.174803} {\bibfield  {journal}
  {\bibinfo  {journal} {Physical Review Letters}\ }\textbf {\bibinfo {volume}
  {128}},\ \bibinfo {pages} {174803} (\bibinfo {year} {2022})}\BibitemShut
  {NoStop}%
\bibitem [{\citenamefont {PPompili}\ \emph {et~al.}(2016)\citenamefont
  {PPompili}, \citenamefont {Anania}, \citenamefont {Bellaveglia},
  \citenamefont {Biagioni}, \citenamefont {Bisesto}, \citenamefont {Chiadroni},
  \citenamefont {Cianchi}, \citenamefont {Croia}, \citenamefont {Curcio},
  \citenamefont {Di~Giovenale}, \citenamefont {Ferrario}, \citenamefont
  {Filippi}, \citenamefont {Galletti}, \citenamefont {Gallo}, \citenamefont
  {Giribono}, \citenamefont {Li}, \citenamefont {Marocchino}, \citenamefont
  {Mostacci}, \citenamefont {Petrarca}, \citenamefont {Petrillo}, \citenamefont
  {Di~Pirro}, \citenamefont {Romeo}, \citenamefont {Rossi}, \citenamefont
  {Scifo}, \citenamefont {Shpakov}, \citenamefont {Vaccarezza}, \citenamefont
  {Villa},\ and\ \citenamefont {Zhu}}]{pompili_beam_2016}%
  \BibitemOpen
  \bibfield  {author} {\bibinfo {author} {\bibfnamefont {R.}~\bibnamefont
  {PPompili}}, \bibinfo {author} {\bibfnamefont {M.}~\bibnamefont {Anania}},
  \bibinfo {author} {\bibfnamefont {M.}~\bibnamefont {Bellaveglia}}, \bibinfo
  {author} {\bibfnamefont {A.}~\bibnamefont {Biagioni}}, \bibinfo {author}
  {\bibfnamefont {F.}~\bibnamefont {Bisesto}}, \bibinfo {author} {\bibfnamefont
  {E.}~\bibnamefont {Chiadroni}}, \bibinfo {author} {\bibfnamefont
  {A.}~\bibnamefont {Cianchi}}, \bibinfo {author} {\bibfnamefont
  {M.}~\bibnamefont {Croia}}, \bibinfo {author} {\bibfnamefont
  {A.}~\bibnamefont {Curcio}}, \bibinfo {author} {\bibfnamefont
  {D.}~\bibnamefont {Di~Giovenale}}, \bibinfo {author} {\bibfnamefont
  {M.}~\bibnamefont {Ferrario}}, \bibinfo {author} {\bibfnamefont
  {F.}~\bibnamefont {Filippi}}, \bibinfo {author} {\bibfnamefont
  {M.}~\bibnamefont {Galletti}}, \bibinfo {author} {\bibfnamefont
  {A.}~\bibnamefont {Gallo}}, \bibinfo {author} {\bibfnamefont
  {A.}~\bibnamefont {Giribono}}, \bibinfo {author} {\bibfnamefont
  {W.}~\bibnamefont {Li}}, \bibinfo {author} {\bibfnamefont {A.}~\bibnamefont
  {Marocchino}}, \bibinfo {author} {\bibfnamefont {A.}~\bibnamefont
  {Mostacci}}, \bibinfo {author} {\bibfnamefont {M.}~\bibnamefont {Petrarca}},
  \bibinfo {author} {\bibfnamefont {V.}~\bibnamefont {Petrillo}}, \bibinfo
  {author} {\bibfnamefont {G.}~\bibnamefont {Di~Pirro}}, \bibinfo {author}
  {\bibfnamefont {S.}~\bibnamefont {Romeo}}, \bibinfo {author} {\bibfnamefont
  {A.}~\bibnamefont {Rossi}}, \bibinfo {author} {\bibfnamefont
  {J.}~\bibnamefont {Scifo}}, \bibinfo {author} {\bibfnamefont
  {V.}~\bibnamefont {Shpakov}}, \bibinfo {author} {\bibfnamefont
  {C.}~\bibnamefont {Vaccarezza}}, \bibinfo {author} {\bibfnamefont
  {F.}~\bibnamefont {Villa}}, \ and\ \bibinfo {author} {\bibfnamefont
  {J.}~\bibnamefont {Zhu}},\ }\href {\doibase 10.1016/j.nima.2016.01.061}
  {\bibfield  {journal} {\bibinfo  {journal} {Nuclear Instruments and Methods
  in Physics Research Section A: Accelerators, Spectrometers, Detectors and
  Associated Equipment}\ }\textbf {\bibinfo {volume} {829}},\ \bibinfo {pages}
  {17} (\bibinfo {year} {2016})}\BibitemShut {NoStop}%
\bibitem [{\citenamefont {Xu}\ \emph {et~al.}(2017)\citenamefont {Xu},
  \citenamefont {Li}, \citenamefont {An}, \citenamefont {Dalichaouch},
  \citenamefont {Yu}, \citenamefont {Lu}, \citenamefont {Joshi},\ and\
  \citenamefont {Mori}}]{xu_high_2017}%
  \BibitemOpen
  \bibfield  {author} {\bibinfo {author} {\bibfnamefont {X.}~\bibnamefont
  {Xu}}, \bibinfo {author} {\bibfnamefont {F.}~\bibnamefont {Li}}, \bibinfo
  {author} {\bibfnamefont {W.}~\bibnamefont {An}}, \bibinfo {author}
  {\bibfnamefont {T.}~\bibnamefont {Dalichaouch}}, \bibinfo {author}
  {\bibfnamefont {P.}~\bibnamefont {Yu}}, \bibinfo {author} {\bibfnamefont
  {W.}~\bibnamefont {Lu}}, \bibinfo {author} {\bibfnamefont {C.}~\bibnamefont
  {Joshi}}, \ and\ \bibinfo {author} {\bibfnamefont {W.}~\bibnamefont {Mori}},\
  }\href {\doibase 10.1103/PhysRevAccelBeams.20.111303} {\bibfield  {journal}
  {\bibinfo  {journal} {Physical Review Accelerators and Beams}\ }\textbf
  {\bibinfo {volume} {20}},\ \bibinfo {pages} {111303} (\bibinfo {year}
  {2017})}\BibitemShut {NoStop}%
\bibitem [{\citenamefont {Liang}\ \emph {et~al.}(2022)\citenamefont {Liang},
  \citenamefont {Xia}, \citenamefont {Pukhov},\ and\ \citenamefont
  {Farmer}}]{liangAccelerationElectronBunch2022}%
  \BibitemOpen
  \bibfield  {author} {\bibinfo {author} {\bibfnamefont {L.}~\bibnamefont
  {Liang}}, \bibinfo {author} {\bibfnamefont {G.}~\bibnamefont {Xia}}, \bibinfo
  {author} {\bibfnamefont {A.}~\bibnamefont {Pukhov}}, \ and\ \bibinfo {author}
  {\bibfnamefont {J.~P.}\ \bibnamefont {Farmer}},\ }\href {\doibase
  10.3390/app122110919} {\bibfield  {journal} {\bibinfo  {journal} {Applied
  Sciences}\ }\textbf {\bibinfo {volume} {12}},\ \bibinfo {pages} {10919}
  (\bibinfo {year} {2022})}\BibitemShut {NoStop}%
\bibitem [{\citenamefont {Si}\ \emph {et~al.}(2023)\citenamefont {Si},
  \citenamefont {Huang}, \citenamefont {Ruan}, \citenamefont {Shen},
  \citenamefont {Xu}, \citenamefont {Yu}, \citenamefont {Wang},\ and\
  \citenamefont {Chen}}]{si_relativistic-guided_2023}%
  \BibitemOpen
  \bibfield  {author} {\bibinfo {author} {\bibfnamefont {M.}~\bibnamefont
  {Si}}, \bibinfo {author} {\bibfnamefont {Y.}~\bibnamefont {Huang}}, \bibinfo
  {author} {\bibfnamefont {M.}~\bibnamefont {Ruan}}, \bibinfo {author}
  {\bibfnamefont {B.}~\bibnamefont {Shen}}, \bibinfo {author} {\bibfnamefont
  {Z.}~\bibnamefont {Xu}}, \bibinfo {author} {\bibfnamefont {T.}~\bibnamefont
  {Yu}}, \bibinfo {author} {\bibfnamefont {X.}~\bibnamefont {Wang}}, \ and\
  \bibinfo {author} {\bibfnamefont {Y.}~\bibnamefont {Chen}},\ }\href {\doibase
  10.1364/OE.503814} {\bibfield  {journal} {\bibinfo  {journal} {Optics
  Express}\ }\textbf {\bibinfo {volume} {31}},\ \bibinfo {pages} {40202}
  (\bibinfo {year} {2023})}\BibitemShut {NoStop}%
\bibitem [{\citenamefont {Si}\ \emph {et~al.}(2024)\citenamefont {Si},
  \citenamefont {Huang}, \citenamefont {Ruan}, \citenamefont {Shen},
  \citenamefont {Xu}, \citenamefont {Yu}, \citenamefont {Wang},\ and\
  \citenamefont {Chen}}]{si_stable_2024}%
  \BibitemOpen
  \bibfield  {author} {\bibinfo {author} {\bibfnamefont {M.}~\bibnamefont
  {Si}}, \bibinfo {author} {\bibfnamefont {Y.}~\bibnamefont {Huang}}, \bibinfo
  {author} {\bibfnamefont {M.}~\bibnamefont {Ruan}}, \bibinfo {author}
  {\bibfnamefont {B.}~\bibnamefont {Shen}}, \bibinfo {author} {\bibfnamefont
  {Z.}~\bibnamefont {Xu}}, \bibinfo {author} {\bibfnamefont {T.}~\bibnamefont
  {Yu}}, \bibinfo {author} {\bibfnamefont {X.}~\bibnamefont {Wang}}, \ and\
  \bibinfo {author} {\bibfnamefont {Y.}~\bibnamefont {Chen}},\ }\href
  {http://arxiv.org/abs/2302.12418} {\enquote {\bibinfo {title} {Stable
  radiation field positron acceleration in a micro-tube},}\ } (\bibinfo {year}
  {2024}),\ \bibinfo {note} {arXiv:2302.12418 [physics]}\BibitemShut {NoStop}%
\bibitem [{\citenamefont {Farmer}\ and\ \citenamefont
  {Porta}(2024)}]{farmerWakefieldRegenerationPlasma2024}%
  \BibitemOpen
  \bibfield  {author} {\bibinfo {author} {\bibfnamefont {J.~P.}\ \bibnamefont
  {Farmer}}\ and\ \bibinfo {author} {\bibfnamefont {G.~Z.~D.}\ \bibnamefont
  {Porta}},\ }\href {http://arxiv.org/abs/2404.14175} {\enquote {\bibinfo
  {title} {Wakefield regeneration in a plasma accelerator},}\ } (\bibinfo
  {year} {2024})\BibitemShut {NoStop}%
\bibitem [{\citenamefont {Qu}\ \emph {et~al.}(2021)\citenamefont {Qu},
  \citenamefont {Meuren},\ and\ \citenamefont
  {Fi}}]{quSignatureCollectivePlasma2021}%
  \BibitemOpen
  \bibfield  {author} {\bibinfo {author} {\bibfnamefont {K.}~\bibnamefont
  {Qu}}, \bibinfo {author} {\bibfnamefont {S.}~\bibnamefont {Meuren}}, \ and\
  \bibinfo {author} {\bibnamefont {Fi}},\ }\href {\doibase
  10.1103/PhysRevLett.127.095001} {\bibfield  {journal} {\bibinfo  {journal}
  {Physical Review Letters}\ }\textbf {\bibinfo {volume} {127}},\ \bibinfo
  {pages} {095001} (\bibinfo {year} {2021})}\BibitemShut {NoStop}%
\bibitem [{\citenamefont {Yakimenko}\ \emph {et~al.}(2019)\citenamefont
  {Yakimenko}, \citenamefont {Alsberg}, \citenamefont {Bong}, \citenamefont
  {Bouchard}, \citenamefont {Clarke}, \citenamefont {Emma}, \citenamefont
  {Green}, \citenamefont {Hast}, \citenamefont {Hogan}, \citenamefont
  {Seabury}, \citenamefont {Lipkowitz}, \citenamefont {O’Shea}, \citenamefont
  {Storey}, \citenamefont {White},\ and\ \citenamefont
  {Yocky}}]{yakimenko_facet-ii_2019}%
  \BibitemOpen
  \bibfield  {author} {\bibinfo {author} {\bibfnamefont {V.}~\bibnamefont
  {Yakimenko}}, \bibinfo {author} {\bibfnamefont {L.}~\bibnamefont {Alsberg}},
  \bibinfo {author} {\bibfnamefont {E.}~\bibnamefont {Bong}}, \bibinfo {author}
  {\bibfnamefont {G.}~\bibnamefont {Bouchard}}, \bibinfo {author}
  {\bibfnamefont {C.}~\bibnamefont {Clarke}}, \bibinfo {author} {\bibfnamefont
  {C.}~\bibnamefont {Emma}}, \bibinfo {author} {\bibfnamefont {S.}~\bibnamefont
  {Green}}, \bibinfo {author} {\bibfnamefont {C.}~\bibnamefont {Hast}},
  \bibinfo {author} {\bibfnamefont {M.}~\bibnamefont {Hogan}}, \bibinfo
  {author} {\bibfnamefont {J.}~\bibnamefont {Seabury}}, \bibinfo {author}
  {\bibfnamefont {N.}~\bibnamefont {Lipkowitz}}, \bibinfo {author}
  {\bibfnamefont {B.}~\bibnamefont {O’Shea}}, \bibinfo {author}
  {\bibfnamefont {D.}~\bibnamefont {Storey}}, \bibinfo {author} {\bibfnamefont
  {G.}~\bibnamefont {White}}, \ and\ \bibinfo {author} {\bibfnamefont
  {G.}~\bibnamefont {Yocky}},\ }\href {\doibase
  10.1103/PhysRevAccelBeams.22.101301} {\bibfield  {journal} {\bibinfo
  {journal} {Physical Review Accelerators and Beams}\ }\textbf {\bibinfo
  {volume} {22}},\ \bibinfo {pages} {101301} (\bibinfo {year}
  {2019})}\BibitemShut {NoStop}%
\bibitem [{\citenamefont {Arber}\ \emph {et~al.}(2015)\citenamefont {Arber},
  \citenamefont {Bennett}, \citenamefont {Brady}, \citenamefont
  {Lawrence-Douglas}, \citenamefont {Ramsay}, \citenamefont {Sircombe},
  \citenamefont {Gillies}, \citenamefont {Evans}, \citenamefont {Schmitz},
  \citenamefont {Bell},\ and\ \citenamefont
  {Ridgers}}]{arber_contemporary_2015}%
  \BibitemOpen
  \bibfield  {author} {\bibinfo {author} {\bibfnamefont {T.~D.}\ \bibnamefont
  {Arber}}, \bibinfo {author} {\bibfnamefont {K.}~\bibnamefont {Bennett}},
  \bibinfo {author} {\bibfnamefont {C.~S.}\ \bibnamefont {Brady}}, \bibinfo
  {author} {\bibfnamefont {A.}~\bibnamefont {Lawrence-Douglas}}, \bibinfo
  {author} {\bibfnamefont {M.~G.}\ \bibnamefont {Ramsay}}, \bibinfo {author}
  {\bibfnamefont {N.~J.}\ \bibnamefont {Sircombe}}, \bibinfo {author}
  {\bibfnamefont {P.}~\bibnamefont {Gillies}}, \bibinfo {author} {\bibfnamefont
  {R.~G.}\ \bibnamefont {Evans}}, \bibinfo {author} {\bibfnamefont
  {H.}~\bibnamefont {Schmitz}}, \bibinfo {author} {\bibfnamefont {A.~R.}\
  \bibnamefont {Bell}}, \ and\ \bibinfo {author} {\bibfnamefont {C.~P.}\
  \bibnamefont {Ridgers}},\ }\href {\doibase 10.1088/0741-3335/57/11/113001}
  {\bibfield  {journal} {\bibinfo  {journal} {Plasma Physics and Controlled
  Fusion}\ }\textbf {\bibinfo {volume} {57}},\ \bibinfo {pages} {113001}
  (\bibinfo {year} {2015})}\BibitemShut {NoStop}%
\bibitem [{\citenamefont {Clarke}\ \emph {et~al.}(2022)\citenamefont {Clarke},
  \citenamefont {Esarey}, \citenamefont {Geddes}, \citenamefont {Hofstaetter},
  \citenamefont {Hogan}, \citenamefont {Nagaitsev}, \citenamefont {Palmer},
  \citenamefont {Piot}, \citenamefont {Power}, \citenamefont {Schroeder},
  \citenamefont {Umstadter}, \citenamefont {Vafaei-Najafabadi}, \citenamefont
  {Valishev}, \citenamefont {Willingale},\ and\ \citenamefont
  {Yakimenko}}]{clarkeUSAdvancedNovel2022}%
  \BibitemOpen
  \bibfield  {author} {\bibinfo {author} {\bibfnamefont {C.}~\bibnamefont
  {Clarke}}, \bibinfo {author} {\bibfnamefont {E.}~\bibnamefont {Esarey}},
  \bibinfo {author} {\bibfnamefont {C.}~\bibnamefont {Geddes}}, \bibinfo
  {author} {\bibfnamefont {G.}~\bibnamefont {Hofstaetter}}, \bibinfo {author}
  {\bibfnamefont {M.}~\bibnamefont {Hogan}}, \bibinfo {author} {\bibfnamefont
  {S.}~\bibnamefont {Nagaitsev}}, \bibinfo {author} {\bibfnamefont
  {M.}~\bibnamefont {Palmer}}, \bibinfo {author} {\bibfnamefont
  {P.}~\bibnamefont {Piot}}, \bibinfo {author} {\bibfnamefont {J.}~\bibnamefont
  {Power}}, \bibinfo {author} {\bibfnamefont {C.}~\bibnamefont {Schroeder}},
  \bibinfo {author} {\bibfnamefont {D.}~\bibnamefont {Umstadter}}, \bibinfo
  {author} {\bibfnamefont {N.}~\bibnamefont {Vafaei-Najafabadi}}, \bibinfo
  {author} {\bibfnamefont {A.}~\bibnamefont {Valishev}}, \bibinfo {author}
  {\bibfnamefont {L.}~\bibnamefont {Willingale}}, \ and\ \bibinfo {author}
  {\bibfnamefont {V.}~\bibnamefont {Yakimenko}},\ }\href {\doibase
  10.1088/1748-0221/17/05/T05009} {\bibfield  {journal} {\bibinfo  {journal}
  {Journal of Instrumentation}\ }\textbf {\bibinfo {volume} {17}},\ \bibinfo
  {pages} {T05009} (\bibinfo {year} {2022})}\BibitemShut {NoStop}%
\bibitem [{\citenamefont {Östling}\ \emph {et~al.}(1997)\citenamefont
  {Östling}, \citenamefont {Tománek},\ and\ \citenamefont
  {Rosén}}]{ostling_electronic_1997}%
  \BibitemOpen
  \bibfield  {author} {\bibinfo {author} {\bibfnamefont {D.}~\bibnamefont
  {Östling}}, \bibinfo {author} {\bibfnamefont {D.}~\bibnamefont {Tománek}},
  \ and\ \bibinfo {author} {\bibfnamefont {A.}~\bibnamefont {Rosén}},\ }\href
  {\doibase 10.1103/PhysRevB.55.13980} {\bibfield  {journal} {\bibinfo
  {journal} {Physical Review B}\ }\textbf {\bibinfo {volume} {55}},\ \bibinfo
  {pages} {13980} (\bibinfo {year} {1997})}\BibitemShut {NoStop}%
\bibitem [{\citenamefont {Wang}\ and\ \citenamefont
  {Mišković}(2004)}]{wang_interactions_2004}%
  \BibitemOpen
  \bibfield  {author} {\bibinfo {author} {\bibfnamefont {Y.-N.}\ \bibnamefont
  {Wang}}\ and\ \bibinfo {author} {\bibfnamefont {Z.~L.}\ \bibnamefont
  {Mišković}},\ }\href {\doibase 10.1103/PhysRevA.69.022901} {\bibfield
  {journal} {\bibinfo  {journal} {Physical Review A}\ }\textbf {\bibinfo
  {volume} {69}},\ \bibinfo {pages} {022901} (\bibinfo {year}
  {2004})}\BibitemShut {NoStop}%
\bibitem [{\citenamefont {Martín-Luna}\ \emph {et~al.}(2023)\citenamefont
  {Martín-Luna}, \citenamefont {Bonatto}, \citenamefont {Bontoiu},
  \citenamefont {Xia},\ and\ \citenamefont
  {Resta-López}}]{martin-luna_excitation_2023}%
  \BibitemOpen
  \bibfield  {author} {\bibinfo {author} {\bibfnamefont {P.}~\bibnamefont
  {Martín-Luna}}, \bibinfo {author} {\bibfnamefont {A.}~\bibnamefont
  {Bonatto}}, \bibinfo {author} {\bibfnamefont {C.}~\bibnamefont {Bontoiu}},
  \bibinfo {author} {\bibfnamefont {G.}~\bibnamefont {Xia}}, \ and\ \bibinfo
  {author} {\bibfnamefont {J.}~\bibnamefont {Resta-López}},\ }\href {\doibase
  10.1088/1367-2630/ad127c} {\bibfield  {journal} {\bibinfo  {journal} {New
  Journal of Physics}\ }\textbf {\bibinfo {volume} {25}},\ \bibinfo {pages}
  {123029} (\bibinfo {year} {2023})}\BibitemShut {NoStop}%
\bibitem [{\citenamefont {Ospina-Bohórquez}\ \emph {et~al.}(2023)\citenamefont
  {Ospina-Bohórquez}, \citenamefont {Salgado-López}, \citenamefont {Ehret},
  \citenamefont {Malko}, \citenamefont {Salvadori}, \citenamefont {Pisarczyk},
  \citenamefont {Chodukowski}, \citenamefont {Rusiniak}, \citenamefont
  {Krupka}, \citenamefont {Lendrin}, \citenamefont {Pérez-Callejo},
  \citenamefont {Vlachos}, \citenamefont {Hannachi}, \citenamefont {Tarisien},
  \citenamefont {Consoli}, \citenamefont {Verona}, \citenamefont {Prestopino},
  \citenamefont {Dostal}, \citenamefont {Dudzak}, \citenamefont {Henares},
  \citenamefont {Apiñaniz}, \citenamefont {DeLuis}, \citenamefont {Debayle},
  \citenamefont {Caron}, \citenamefont {Ceccotti}, \citenamefont
  {Hernández-Martín}, \citenamefont {Hernández-Toro}, \citenamefont
  {Huault}, \citenamefont {Martín-López}, \citenamefont {Méndez},
  \citenamefont {Nguyen-Bui}, \citenamefont {Perez-Hernández}, \citenamefont
  {Vaisseau}, \citenamefont {Varela}, \citenamefont {Volpe}, \citenamefont
  {Gremillet},\ and\ \citenamefont
  {Santos}}]{ospina-bohorquez_laser-driven_2023}%
  \BibitemOpen
  \bibfield  {author} {\bibinfo {author} {\bibfnamefont {V.}~\bibnamefont
  {Ospina-Bohórquez}}, \bibinfo {author} {\bibfnamefont {C.}~\bibnamefont
  {Salgado-López}}, \bibinfo {author} {\bibfnamefont {M.}~\bibnamefont
  {Ehret}}, \bibinfo {author} {\bibfnamefont {S.}~\bibnamefont {Malko}},
  \bibinfo {author} {\bibfnamefont {M.}~\bibnamefont {Salvadori}}, \bibinfo
  {author} {\bibfnamefont {T.}~\bibnamefont {Pisarczyk}}, \bibinfo {author}
  {\bibfnamefont {T.}~\bibnamefont {Chodukowski}}, \bibinfo {author}
  {\bibfnamefont {Z.}~\bibnamefont {Rusiniak}}, \bibinfo {author}
  {\bibfnamefont {M.}~\bibnamefont {Krupka}}, \bibinfo {author} {\bibfnamefont
  {P.~G.}\ \bibnamefont {Lendrin}}, \bibinfo {author} {\bibfnamefont
  {G.}~\bibnamefont {Pérez-Callejo}}, \bibinfo {author} {\bibfnamefont
  {C.}~\bibnamefont {Vlachos}}, \bibinfo {author} {\bibfnamefont
  {F.}~\bibnamefont {Hannachi}}, \bibinfo {author} {\bibfnamefont
  {M.}~\bibnamefont {Tarisien}}, \bibinfo {author} {\bibfnamefont
  {F.}~\bibnamefont {Consoli}}, \bibinfo {author} {\bibfnamefont
  {C.}~\bibnamefont {Verona}}, \bibinfo {author} {\bibfnamefont
  {G.}~\bibnamefont {Prestopino}}, \bibinfo {author} {\bibfnamefont
  {J.}~\bibnamefont {Dostal}}, \bibinfo {author} {\bibfnamefont
  {R.}~\bibnamefont {Dudzak}}, \bibinfo {author} {\bibfnamefont {J.~L.}\
  \bibnamefont {Henares}}, \bibinfo {author} {\bibfnamefont {J.~I.}\
  \bibnamefont {Apiñaniz}}, \bibinfo {author} {\bibfnamefont {D.}~\bibnamefont
  {DeLuis}}, \bibinfo {author} {\bibfnamefont {A.}~\bibnamefont {Debayle}},
  \bibinfo {author} {\bibfnamefont {J.}~\bibnamefont {Caron}}, \bibinfo
  {author} {\bibfnamefont {T.}~\bibnamefont {Ceccotti}}, \bibinfo {author}
  {\bibfnamefont {R.}~\bibnamefont {Hernández-Martín}}, \bibinfo {author}
  {\bibfnamefont {J.}~\bibnamefont {Hernández-Toro}}, \bibinfo {author}
  {\bibfnamefont {M.}~\bibnamefont {Huault}}, \bibinfo {author} {\bibfnamefont
  {A.}~\bibnamefont {Martín-López}}, \bibinfo {author} {\bibfnamefont
  {C.}~\bibnamefont {Méndez}}, \bibinfo {author} {\bibfnamefont {T.-H.}\
  \bibnamefont {Nguyen-Bui}}, \bibinfo {author} {\bibfnamefont {J.~A.}\
  \bibnamefont {Perez-Hernández}}, \bibinfo {author} {\bibfnamefont
  {X.}~\bibnamefont {Vaisseau}}, \bibinfo {author} {\bibfnamefont
  {O.}~\bibnamefont {Varela}}, \bibinfo {author} {\bibfnamefont
  {L.}~\bibnamefont {Volpe}}, \bibinfo {author} {\bibfnamefont
  {L.}~\bibnamefont {Gremillet}}, \ and\ \bibinfo {author} {\bibfnamefont
  {J.~J.}\ \bibnamefont {Santos}},\ }\href {http://arxiv.org/abs/2311.04187}
  {\enquote {\bibinfo {title} {Laser-driven ion and electron acceleration from
  near-critical density gas targets: towards high-repetition rate operation in
  the 1 {PW}, sub-100 fs laser interaction regime},}\ } (\bibinfo {year}
  {2023}),\ \bibinfo {note} {arXiv:2311.04187 [physics]}\BibitemShut {NoStop}%
\bibitem [{Jac(1998)}]{Jackson1998}%
  \BibitemOpen
  \href@noop {} {\emph {\bibinfo {title} {Classical Electrodynamics}}},\
  \bibinfo {edition} {3rd}\ ed.,\ Hardcover 832 pages\ (\bibinfo  {publisher}
  {John Wiley \& Sons},\ \bibinfo {year} {1998})\BibitemShut {NoStop}%
\bibitem [{\citenamefont {Merrill}\ \emph {et~al.}(2018)\citenamefont
  {Merrill}, \citenamefont {Goett}, \citenamefont {Goett}, \citenamefont
  {Gibbs}, \citenamefont {Imhoff}, \citenamefont {Mariam}, \citenamefont
  {Morris}, \citenamefont {Neukirch}, \citenamefont {Perry}, \citenamefont
  {Poulson}, \citenamefont {Simpson}, \citenamefont {Volegov}, \citenamefont
  {Walstrom}, \citenamefont {Wilde}, \citenamefont {Hast}, \citenamefont
  {Jobe}, \citenamefont {Smith}, \citenamefont {Wienands}, \citenamefont
  {Clarke},\ and\ \citenamefont {Tourret}}]{merrill_demonstration_2018}%
  \BibitemOpen
  \bibfield  {author} {\bibinfo {author} {\bibfnamefont {F.~E.}\ \bibnamefont
  {Merrill}}, \bibinfo {author} {\bibfnamefont {F.~E.}\ \bibnamefont {Goett},
  \bibfnamefont {JMerrill}}, \bibinfo {author} {\bibfnamefont {J.}~\bibnamefont
  {Goett}}, \bibinfo {author} {\bibfnamefont {J.~W.}\ \bibnamefont {Gibbs}},
  \bibinfo {author} {\bibfnamefont {S.~D.}\ \bibnamefont {Imhoff}}, \bibinfo
  {author} {\bibfnamefont {F.~G.}\ \bibnamefont {Mariam}}, \bibinfo {author}
  {\bibfnamefont {C.~L.}\ \bibnamefont {Morris}}, \bibinfo {author}
  {\bibfnamefont {L.~P.}\ \bibnamefont {Neukirch}}, \bibinfo {author}
  {\bibfnamefont {J.}~\bibnamefont {Perry}}, \bibinfo {author} {\bibfnamefont
  {D.}~\bibnamefont {Poulson}}, \bibinfo {author} {\bibfnamefont
  {R.}~\bibnamefont {Simpson}}, \bibinfo {author} {\bibfnamefont {P.~L.}\
  \bibnamefont {Volegov}}, \bibinfo {author} {\bibfnamefont {P.~L.}\
  \bibnamefont {Walstrom}}, \bibinfo {author} {\bibfnamefont {C.~H.}\
  \bibnamefont {Wilde}}, \bibinfo {author} {\bibfnamefont {C.}~\bibnamefont
  {Hast}}, \bibinfo {author} {\bibfnamefont {K.}~\bibnamefont {Jobe}}, \bibinfo
  {author} {\bibfnamefont {T.}~\bibnamefont {Smith}}, \bibinfo {author}
  {\bibfnamefont {U.}~\bibnamefont {Wienands}}, \bibinfo {author}
  {\bibfnamefont {A.~J.}\ \bibnamefont {Clarke}}, \ and\ \bibinfo {author}
  {\bibfnamefont {D.}~\bibnamefont {Tourret}},\ }\href {\doibase
  10.1063/1.5011198} {\bibfield  {journal} {\bibinfo  {journal} {Applied
  Physics Letters}\ }\textbf {\bibinfo {volume} {112}},\ \bibinfo {pages}
  {144103} (\bibinfo {year} {2018})}\BibitemShut {NoStop}%
\bibitem [{\citenamefont {{Walstrom, Peter and Barber, Ronald and Chapman,
  Catherine and Garnett, Robert and Gomez, Tony and O'Toole, Joseph and
  Salazar, Harry}}(2015)}]{Walstrom:2015bbt}%
  \BibitemOpen
  \bibfield  {author} {\bibinfo {author} {\bibnamefont {{Walstrom, Peter and
  Barber, Ronald and Chapman, Catherine and Garnett, Robert and Gomez, Tony and
  O'Toole, Joseph and Salazar, Harry}}},\ }in\ \href {\doibase
  10.18429/JACoW-IPAC2015-TUPWI027} {\emph {\bibinfo {booktitle} {{6th
  International Particle Accelerator Conference}}}}\ (\bibinfo {year} {2015})\
  p.\ \bibinfo {pages} {TUPWI027}\BibitemShut {NoStop}%
\bibitem [{\citenamefont {Gai}\ \emph {et~al.}(2014)\citenamefont {Gai},
  \citenamefont {Qiu},\ and\ \citenamefont
  {Jing}}]{bell_electron_2014_10.1117/12.2061952}%
  \BibitemOpen
  \bibfield  {author} {\bibinfo {author} {\bibfnamefont {W.}~\bibnamefont
  {Gai}}, \bibinfo {author} {\bibfnamefont {J.}~\bibnamefont {Qiu}}, \ and\
  \bibinfo {author} {\bibfnamefont {C.}~\bibnamefont {Jing}},\ }in\ \href
  {\doibase 10.1117/12.2061952} {\emph {\bibinfo {booktitle} {Target
  Diagnostics Physics and Engineering for Inertial Confinement Fusion III}}},\
  Vol.\ \bibinfo {volume} {9211},\ \bibinfo {editor} {edited by\ \bibinfo
  {editor} {\bibfnamefont {P.~M.}\ \bibnamefont {Bell}}\ and\ \bibinfo {editor}
  {\bibfnamefont {G.~P.}\ \bibnamefont {Grim}}},\ \bibinfo {organization}
  {International Society for Optics and Photonics}\ (\bibinfo  {publisher}
  {SPIE},\ \bibinfo {year} {2014})\ p.\ \bibinfo {pages} {921104}\BibitemShut
  {NoStop}%
\end{thebibliography}%

%
	

\end{document}